\pdfoutput=1
\documentclass[3p,twocolumn]{elsarticle}

\usepackage{lineno,hyperref}
\usepackage{afterpage}
\usepackage{dblfloatfix}  
\modulolinenumbers[5]

\usepackage{todonotes}
\usepackage{subcaption}
\usepackage{amsmath}
\usepackage{multirow}

\begin{document}

\begin{frontmatter}

\title{A Non-Linear Kalman Filter for track parameters estimation in High Energy Physics}


\author[address1]{Xiaocong Ai\corref{correspondingauthor}}
\cortext[correspondingauthor]{Corresponding author}
\ead{xiaocong.ai@desy.de}

\author[address2,address3]{Heather M. Gray}\ead{heather.gray@berkeley.edu}
\author[address4]{Andreas Salzburger}\ead{andreas.salzburger@cern.ch}
\author[address1]{Nicholas Styles}\ead{nicholas.styles@desy.de}

\address[address1]{Deutsches Elektronen-Synchrotron DESY, Notkestr. 85, 22607 Hamburg, Germany}
\address[address2]{Department of Physics, University of California, 425 Physics South MC 7300 Berkeley, CA, 94720, USA}
\address[address3]{Physics Division, Lawrence Berkeley National Laboratory, 1 Cyclotron Road, Berkeley, CA 94720, USA}
\address[address4]{CERN, Espl. des Particules 1, 1217 Meyrin, Switzerland}

\begin{abstract}
The Kalman Filter is a widely used approach for the linear estimation of dynamical systems and is frequently employed within nuclear and particle physics experiments for the reconstruction of charged particle trajectories, known as tracks. Implementations of this formalism often make assumptions on the linearity of the underlying dynamic system and the Gaussian nature of the process noise, which is violated in many track reconstruction applications. This paper introduces an implementation of a Non-Linear Kalman Filter (NLKF) within the ACTS track reconstruction toolkit. The NLKF addresses the issue of non-linearity by using a set of representative sample points during its track state propagation. In a typical use case, the NLKF outperforms an Extended Kalman Filter in the accuracy and precision of the track parameter estimates obtained, with the increase in CPU time below a factor of two. It is therefore a promising approach for use in applications where precise estimation of track parameters is a key concern. 
\end{abstract}

\begin{keyword}
\texttt Non-linear system \sep Non-linear Kalman filter \sep Nuclear and particle physics experiment \sep Track parameter estimates
\MSC[2021] 00-01\sep  99-00
\end{keyword}

\end{frontmatter}


\section{Introduction}

The reconstruction of the trajectories of charged particles requires the identification of the set of hits corresponding to a single particle and the determination of the kinematic properties of the particle's trajectory by fitting that set of hits.
The most commonly used algorithm for the reconstruction of charged particle trajectories, or tracks, in nuclear and particle physics is the Kalman Filter (KF). The KF was introduced approximately 70 years ago~\cite{kalman1960} and is used in many fields including navigation, aerospace engineering, space engineering, remote surveillance, telecommunications, physics, audio signal processing and control engineering.

The KF processes a set of discrete measurements to determine the internal state of a linear dynamical system. Both the measurements and the system can be subjected to independent random perturbations or noise. By combining predictions based on the previous state estimates with subsequent measurements, the impact of these perturbations on the following state estimates can be minimized. The Kalman filter is known to be the optimal linear estimator for such linear systems.

The KF for track reconstruction was introduced to particle physics by the DELPHI experiment~\cite{DELPHI:1995dsm} at the Large Electron Positron (LEP) collider at the European Council for Nuclear Research (CERN). In track reconstruction~\cite{Fruhwirth:1987fm}, the description of the system incorporates the impact of magnetic fields and detector material on charged particle trajectories\footnote{Magnetic fields are used to deflect the trajectory to allow the charged particle momentum to be measured and material effects cause random fluctuations due to elastic scattering and energy loss}. KF algorithms are used both in track finding, where the collection of measurements corresponding to a single charged particle trajectory are identified, and in track fitting, where the parameters describing the trajectory of the charged particle are determined from a set of measurements. To date, the KF remains the method with the best overall performance for most track reconstruction applications. See Ref~\cite{RevModPhys.82.1419} for a review.

KF algorithms for track reconstruction typically proceed in two steps. The starting point is the track seed, which is an initial coarse trajectory estimate for a candidate track, based on a small number of measurements, typically three or four. Subsequent measurements are added progressively to the track seed following a track propagation to reachable detection elements. Once all the measurements have been added, a second smoothing step~\cite{RauchTungStriebel} is performed, which runs a second filtering sequence in the opposite direction. This means that information from all measurements are included in the track parameter estimates at all measurement points. Without the smoothing step, only the parameters estimated at the final measurement point would include the information from all measurement points due to the progressive nature of the KF procedure. An extension of the KF is the Combinatorial Kalman Filter (CKF)~\cite{BILLOIR1989390,BILLOIR1990219,MANKEL1997169}, which can be used to perform track finding and track fitting simultaneously and allows branching of track candidates.

Despite the success of the KF, a key limitation for many applications is the assumption of linear models for the system and measurement and Gaussian distributions for the system state, process and measurement noise. This has motivated the development of a number of extensions. One such extension is the Extended Kalman filter (EKF)~\cite{Daum2015} which linearizes a model using a first-order Taylor expansion. This improved description is insufficient in particular when the incidence angle of the charged particle on the measurement surface is large. The EKF assumes that the contribution from the noise is described by a Gaussian distribution, which is not necessarily appropriate.

The Gaussian Sum Filter (GSF)~\cite{FRUHWIRTH199880} relaxes the assumption of Gaussian process noise by assuming that the noise distribution can be described by a sum of Gaussian distributions~\cite{FRUHWIRTH2003131}. In the domain of nuclear and particle physics, this is particularly important when modelling radiative energy loss such as is common when electrons lose energy through bremsstrahlung when passing through tracking detectors~\cite{PERF-2017-01, CMS-EGM-13-001}.
The application of the GSF procedure is typically restricted to track candidates which have been identified as being a potential electron candidate (e.g.~by combining track information with calorimeter information, or other forms of particle identification such as transition radiation). The GSF does not address non-linear effects in tracking fitting. 

This paper will explore a non-linear Kalman filter (NLKF) based on the Unscented Kalman filter (UKF)~\cite{Julier1997NewEO,Julier2004}, which uses a set of discretely sampled points to parameterize the mean and covariance to account for non-linearities of the system and measurement models. It has been shown to have comparable performance to a second-order Gaussian filter. We investigate the application of the UKF to charged particle reconstruction for high-energy nuclear and particle physics experiments.

The manuscript is organized as follows. Section \ref{sec:acts} provides a brief introduction to track reconstruction and A Common Tracking Software Toolkit (ACTS)~\cite{ai2021common}. The formalism for the EKF is discussed in Section~\ref{sec:EKF} and the extension to the NLKF in Section~\ref{sec:NLKF}. Section~\ref{sec:perf} compares the performance of the EKF and the NLKF using a typical detector geometry. Brief conclusions are presented in Section~\ref{sec:concl}.
\label{sec:intro}
\section{Track reconstruction and the ACTS toolkit}
A Common Tracking Software (ACTS) is a toolkit providing a set of encapsulated track reconstruction components that can be used by a wide range of experiments. ACTS features an internal geometry and navigation model, including a minimal Event Data Model (EDM) implementation that allows client applications to augment and extend the data with information specific to the target experiment. It imposes minimal dependencies. ACTS is written in C++17 using modern programming best-practises and follows a component level design that provides encapsulated, stateless modules. These modules perform well-defined tasks for track reconstruction (e.g.~track propagation or track fitting) and are designed to be executed in parallel call paths if desired, in compliance with modern multi-core CPU architectures. ACTS is currently used within a number of nuclear and particle physics experiments, e.g.~ATLAS~\cite{Aad:2008zzm}, sPHENIX~\cite{Osborn2021} and FASER~\cite{Ariga:2019ufm}, and is being investigated as a potential track reconstruction library by a number of others~\cite{abe2010belle,thecepcstudygroup2018cepc_v2,SMYRSKI201285,accardi2014electron,akesson2018light}. 
 
 Based on its internal geometry and navigation model, ACTS provides a fast\footnote{i.e. Fatras is significantly simplified with respect to a physics-based simulation such as Geant4~\cite{Agostinelli:2002hh}, resulting in orders-of-magnitude faster processing times} track simulation engine, based on the concept of the ATLAS Fast Track Simulation (Fatras)~\cite{Edmonds:1091969}. The internal navigation model of the ACTS geometry is used to predict the particle trajectories through the detector. Hits are created at the intersection points of the trajectory with sensitive detector elements, and the interaction of particles with detector material is modelled using approximate electromagnetic and hadronic physics models. The recorded hits are processed by a digitization module that emulates the detector readout and provides an estimate for the detector resolution.

In ACTS, candidate tracks are created from input measurements by a series of track reconstruction algorithms, and are represented by a series of track states, representing the trajectory at various points. A track state can be expressed in either a free (also called global) or a local representation. Local representations are constrained to a surface description within the detector.

The free (global) track parameters, $g$, are 8-dimensional and represented as:
\begin{equation}
 g = (x, y, z, t, d_x, d_y, d_z, q/p).   
 \label{eq:gloablparams}
\end{equation}
The first four parameters are the space-time ($x$, $y$ and $z$ for position and $t$ for time) coordinates of the track state, $d_x$, $d_y$ and $d_z$ represent the direction of the track at that point, and $q/p$ is the ratio of the charge, $q$, and momentum, $p$.
The local track parameters, $l$, are 6-dimensional and represented as\footnote{We assume a right-handed coordinate system, with the polar angle $\theta$ measured from the positive $z$-axis in an interval of $[0,\pi]$, and the azimuthal angle $\phi \in [-\pi,\pi)$ defined in the transverse $x$-$y$ plane, with $\phi=0$ denoting the $x$-axis}:
\begin{equation}
l = (loc_0, loc_1, \phi, \theta, q/p, t).
\end{equation}
Here, $loc_0$ and $loc_1$ are the coordinates of the track in the local coordinate frame of a reference surface, the $\phi$ and $\theta$ are angles representing the track direction in the polar frame, and the $q/p$ and $t$ are the same as in the global track parameters. The reference surface can consist of different types or shapes, including cylindrical or planar surfaces, or surfaces describing straw-like detector or virtual lines. An example of a line surface is the perigee surface used to describe the track parameters near the vertex\footnote{The vertex is assumed to be the common point where particles from a single interaction or decay originated. ACTS also includes algorithms for reconstructing the positions of such vertices from a set of input tracks}. The track parameters on a perigee surface are called the perigee track parameters and, in this case, the $loc_0$ and $loc_1$ are often denoted as $d_0$ and $z_0$, which are the transverse and longitudinal impact parameters. The perigee parameters are often used when the track is described by a single set of parameters at its estimated point of production, which is typically of most relevance for physics analyses. See Ref.~\cite{ai2021common} for more details of the track parametrization.

In the ACTS Kalman filtering algorithm, the track state is represented by the local track parameters expressed at measurement planes. Measurements are represented by a subset of the local track parameters, as explained in Section~\ref{subsec:acts_impl}. 

\label{sec:acts}
\section{Track fitting with Extended Kalman filter}

Track fitting with a Kalman filter requires evolving the track state and its associated covariance matrix, as it is propagated through a discrete dynamical system. This can be described by a track state propagation model:
\begin{equation}
     x_{k} = f_{k-1}(x_{k-1}) + \eta_{k-1}.
     \label{eq:propmodel}
\end{equation}

Here,
\begin{itemize}
    \item $x_{k-1}$ and $x_{k}$ are the track state vector at the states $k-1$ and $k$, respectively.
    \item $\eta_{k-1}$ is the vector representing the noise when propagating from state $k-1$ to state $k$, i.e.~process noise.
    \item  $f_{k-1}$ is the track state propagation model from $k-1$ to state $k$, which describes the motion of the particle. It depends on the kinematics of the particle and the magnetic field. 
\end{itemize}
 
The track state is projected onto the measurement using the measurement projection model:

\begin{equation}
    y_{k} = h_k(x_k) + \epsilon_k.
    \label{eq:measuremodel}
\end{equation}
   
Here, 
\begin{itemize}
    \item $y_{k}$ is the measurement vector at state $k$.
    \item $\epsilon_k$ is the measurement noise vector at state $k$.
    \item $h_{k}$ is the measurement projection function from track state to measurement, which depends on the kinematics of the particles and detector geometry.
\end{itemize}

Both the track state propagation model, $f$, and the measurement projection model, $h$, are often non-linear functions. The process noise $\eta_{k-1}$ and measurement noise $\epsilon_k$ are assumed to be Gaussian distributions with zero means, and variances $Q_k$ and $V_k$ respectively, however they may not necessarily follow Gaussian distributions.

For the EKF, the $f$ and $h$ are approximated with linear models as follows:
\begin{equation}
\renewcommand{\arraystretch}{1.6}
\begin{array}{ll}
     x_{k} = F_{k-1} x_{k-1} + \eta_{k-1}, \\
     y_{k} = H_k x_k + \epsilon_k,
     \label{eq:trackmodel}
\end{array}
\end{equation}
 where $F_{k-1}$ is the first-order Taylor expansion of the track propagation function $f$ at state $k-1$, and $H_k$ is the first-order Taylor expansion of the measurement projection function $h$ at state $k$. As before, $\eta$ and $\epsilon$ are the process and measurement noise vectors.
 
In nuclear and particle experiments which often have inhomogeneous magnetic fields, $F_{k-1}$ is calculated using the Runge-Kutta method~\cite{Myrheim:1979ng} to numerically solve the second-order differential equations describing charged particles moving through magnetic fields. For example, the ATLAS experiment uses an adaptive Runge-Kutta-Nyström approach~\cite{Lund_2009}, which adapts the step size to minimize computational costs while ensuring that the estimation error remains below a set threshold. The $H_k$ matrix is obtained by analytically calculating the derivative of $h$ with respect to the track state vector and accounting for the angle at which the track intersects the detector module.

The EKF includes three steps: the \emph{prediction} of the track state at state $k$ based on previous $k-1$ measurements, the \emph{filtering} of predicted track state at state $k$ taking into account the measurement at state $k$, and the \emph{smoothing} of the filtered track state with all measurements taken into account. A full description can be found in Ref~\cite{Fruhwirth:1987fm}. Here we briefly outline the formulae used to update the track state vector, $x$ and its covariance, $C$.
\begin{itemize}
    \item{Prediction:}
    \begin{equation}
    \renewcommand{\arraystretch}{1.6}
        \begin{array}{ll}
        x_k^{k-1} = F_{k-1}x_{k-1}, \\
        C_k^{k-1} = F_{k-1}C_{k-1}F_{k-1}^T,
        \end{array}
    \end{equation}
    where the upper index $k-1$ indicates the estimate prior to the filtering, i.e.~with only the previous $k-1$ measurements taken into account.
   
    \item{Filtering:}
    \begin{equation}
    \renewcommand{\arraystretch}{1.6}
        \begin{array}{ll}
        x_k = x_{k}^{k-1} + K_k(m_k - H_kx_k^{k-1}), \\
        C_k = (1-K_kH_k)C_k^{k-1},
        \label{eq:filterupdate}
        \end{array}
    \end{equation}
    where $m_k$ is the measurement on state $k$, and the $K_k$ is the Kalman gain matrix:
    \begin{equation}
    K_k = C_k^{k-1}H_k^T(V_k + H_kC_k^{k-1}H_k^T)^{-1}.
    \label{eq:filtergain}
    \end{equation}
  
   \item{Smoothing:}
    \begin{equation}
    \renewcommand{\arraystretch}{1.6}
        \begin{array}{ll}
    x_k^{n} = x_k + A_k(x_{k+1}^n - x_{k+1}^k), \\
    C_k^{n} = C_k + A_k(C_{k+1}^n - C_{k+1}^k)A_k^T,
        \end{array}
    \end{equation}
    where the upper index $n$ indicates the smoothed estimation with all $n$ measurements taken into account, and the $A_k$ is the smoother gain matrix:
     \begin{equation}
     A_k = C_kF_k^T(C_{k+1}^k)^{-1}.
     \end{equation}
\end{itemize}

\label{sec:EKF}
\section{The Non-linear Kalman filter}
\label{sec:NLKF}

\subsection{Non-linear effects in track reconstruction}
\begin{figure}[htb!]
\centering
  \includegraphics[width=1.0\linewidth]{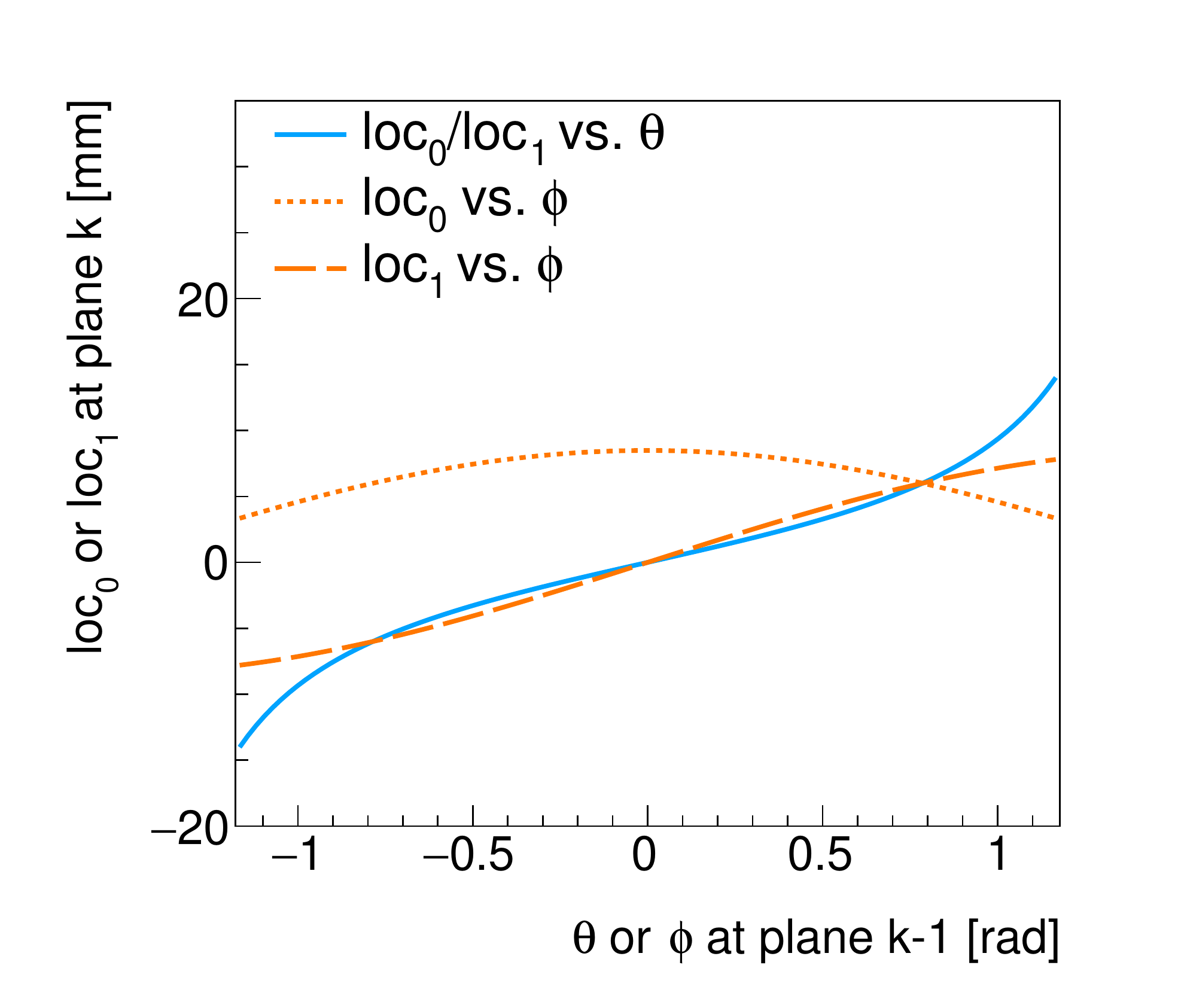}
  \caption{Example of non-linear dependence of the local coordinate of the intersection of a track on detector plane $k$ on the track direction on detector plane $k-1$ for two parallel detector planes oriented normal to the beam direction. The $loc_0$ and $loc_1$ have the same dependence on the $\theta$, which is shown by the solid blue line. The dependence of the $loc_0$ and $loc_1$ on the $\phi$ are shown by the short dashed and long dashed orange lines, respectively.}
  \label{fig:locals_vs_angles}
\end{figure}

Tracking detectors at particle colliders follow a cylindrical or layered approach. A cylindrical detector typically consists of concentric cylindrical layers, which are oriented parallel to the beam direction, in the barrel, and disk layers, which are oriented normal to the beam direction, in the forward regions. This guarantees a close to hermetic coverage of the phase space of the particles produced in the collisions, while complying with mechanical constraints and minimizing detector material. When a track from the beam interaction point intersects with a detector module, the dependence of the intersection position on the incident track direction is non-linear. Fig.~\ref{fig:locals_vs_angles} demonstrates an example of such non-linearity for simplified detector consisting of two parallel detector planes. The local coordinates of the state $k$ are shown as a function of the polar and azimuthal angles of the track direction at the previous state $k-1$. In this example the functions are closest to linear when the azimuthal angle and polar angle are zero, which corresponds to the case when the track intersects the detector module at a perpendicular angle, or zero incidence angle, and become increasingly non-linear when the absolute angles get larger. This effect is particularly significant for the polar angle, which is highly correlated with the track incidence angle. These non-linear effects can be addressed by the NLKF.

\subsection{NLKF formalism}
\label{subsec:NLKFFormalism}
 The NLKF calculates the propagated or projected track state and covariance using a set of sample points around the mean of the track state being propagated or projected, with each point assigned a weight. This is analogous to the random sampling of a distribution function in Monte Carlo simulation, which is the method typically used to generate events corresponding to different physics processes. For a $N$-dimensional track state vector $x_{k-1}$ with covariance $C_{k-1}$ at state $k-1$, $2N+1$ sample points are generated~\cite{Julier1997NewEO,Julier2004}. This comprises the nominal track state vector plus $2N$ vectors generated by varying the nominal track state vector along the direction of the eigenvectors of the covariance matrix. The magnitudes of the variations are given by the eigenvalues of the covariance matrix. 
 The eigenvectors and eigenvalues of the covariance matrix are obtained via Singular Value Decomposition (SVD)~\cite{roth2011}. $C_{k-1}$ is a real symmetric matrix and therefore can be expressed through SVD as,
 
\begin{equation}
    C_{k-1} = U_{k-1}S_{k-1} U_{k-1}^T,
\end{equation}
where $U_{k-1}$ is a unitary matrix whose columns are the eigenvectors of $C_{k-1}$, and $S_{k-1}$ is a diagonal matrix whose non-zero diagonal elements are the corresponding eigenvalues of $C_{k-1}$.
Denoting the $i$-th column of $U_{k-1}$ as $u_{k-1}^{i}$ and the $i$-th diagonal element of $S_{k-1}$ as $s_{k-1}^{i}$, $N$ sets of orthogonal shifting vectors $\delta_{k-1}^{i}$ are,
\begin{equation}
   \delta_{k-1}^{i} = \sqrt{s_{k-1}^{i}} u_{k-1}^{i}, \quad i= 1, ..., N, 
\end{equation}
where $\sqrt{s_{k-1}^{i}}$ is the magnitude of the variation in the direction of $u_{k-1}^{i}$.

The $2N+1$ sample points for $x_{k-1}$ are:

\begin{equation}
\begin{small}
   \textrm{X}_{k-1}^{(i)} = \renewcommand{\arraystretch}{1.6} \left \{
   \begin{array}{ll}
   x_{k-1}, \quad i =0; \\
   x_{k-1} + \gamma \delta_{k-1}^{i},~i= 1, ..., N; \\
   x_{k-1} - \gamma \delta_{k-1}^{i-N},~i= N+1 ,..., 2N,
   \end{array}
  \right.
  \end{small}
\end{equation}

where $\gamma$ is a scaling parameter,
\begin{equation}
\gamma = \sqrt{N+\lambda},~~\lambda = \alpha^2N - N,  
\end{equation}
and $\alpha$ is a tuning parameter used to control the deviation of the sample point from the nominal point, in the range $0 <\alpha \leq 1$.

In principle, the sample points could be propagated using the track model using Eq.~\ref{eq:propmodel},
\begin{equation}
 \textrm{X}_{k}^{(i)} = f_{k-1}(\textrm{X}_{k-1}^{(i)}) + \eta_{k-1}, \quad i = 0, ..., 2N, 
 \label{eq:propmodel_sample}
\end{equation} 
and projected to a measurement point using the measurement model with Eq.~\ref{eq:measuremodel},
\begin{equation}
 \textrm{Y}_k^{(i)} = h_k(\textrm{X}_{k}^{(i)}) + \epsilon_k, \quad i = 0, ..., 2N.
 \label{eq:measuremodel_sample}
\end{equation}
However, because the Runge-Kutta method already accounts for the non-linearity of the track model, we apply only the second of these non-linearity corrections: the projection of the track state to the measurement point.

The mean, $y_k$, and covariance, $P_k$, of the projected track state are calculated as,

\begin{equation}
\renewcommand{\arraystretch}{2.5}
\begin{array}{ll}
y_k = \sum\limits_{i=0}^{2N} w_m^{(i)}\textrm{Y}_{k}^{(i)},\\
P_{k} = \sum\limits_{i=0}^{2N} w_c^{(i)}(\textrm{Y}_{k}^{(i)}-y_k)(\textrm{Y}_{k}^{(i)}-y_k)^T + V_k,
\label{eq:filterEKF}
\end{array}
\end{equation}

and the covariance between the track state and the measurement, $T_k$ is calculated as
\begin{equation}
\begin{array}{ll}
 T_{k} = \sum\limits_{i=0}^{2N}w_c^{(i)}(\textrm{X}_k^{(i)} - x_k^{k-1})(\textrm{Y}_{k}^{(i)}-y_k)^T.
 \end{array}
 \label{eq:crossEKF}
\end{equation}

In Eq.~\ref{eq:filterEKF} and Eq.~\ref{eq:crossEKF}, the weights $w_m^{(i)}$ and $w_c^{(i)}$ are defined as,

\begin{equation}
\renewcommand{\arraystretch}{1.6}
\begin{array}{ll}
w_m^{(0)} = \frac{\lambda}{N+\lambda}, \quad i =0,\\
w_c^{(0)} = \frac{\lambda}{N+\lambda} + (1 -\alpha^2 +\beta), \quad i =0, \\
w_m^{(i)} = w_c^{(i)} = \frac{1}{2(N+\lambda)}, \quad i= 1, ..., 2N,
\end{array}
\end{equation}

where $\beta$ is a non-negative weighting parameter used to tune the weight of the $\textrm{Y}^{(0)}$ when calculating $P_k$. A value of $\beta =2 $ as suggested in Ref.~\cite{Merwe03sigma-pointkalman} is used.

The Kalman gain is calculated as,
\begin{equation}
  K_k = T_{k} P_{k}^{-1},
\end{equation} and used, with the mean and covariance, to update the track state and its covariance in the Kalman filtering step,

\begin{eqnarray}
\renewcommand{\arraystretch}{1.6}
\begin{array}{ll}
x_k = x_{k}^{k-1} + K_k(m_k - y_k), \\
C_k = C_k^{k-1} - K_k P_k K_k^T.
\end{array}
\end{eqnarray}

\subsection{Implementation of NLKF in ACTS}
\label{subsec:acts_impl}
As described in Section~\ref{sec:acts}, a measurement is described by a subset of the local track parameters in ACTS. Therefore, projecting a track state to a measurement is equivalent to transforming the global track parameters to the local track parameters and projecting the local track parameters to the measurement by an identity projection matrix. The track state is represented by global track parameters during its propagation between detector planes and transformed to local track parameters at the detector plane where a material effect needs to be taken into account or a measurement is present. In the latter case, the measurement is used to update the predicted track state $x_k^{k-1}$ and its covariance $C_k^{k-1}$ represented by the local track parameters at state $k$ using Eq.~\ref{eq:filterupdate}. 

If the incidence angle of track on a detector plane is larger than a certain value, the transformation of a single set of global track parameters to local track parameters is replaced by the transformation of the 17 sets\footnote{As discussed in Sec.~\ref{subsec:NLKFFormalism}, the NLKF uses $2N+1$ samples points and $N$ is 8 for the global track parameters} of global track parameters to the local track parameters.  Eq.~\ref{eq:measuremodel_sample} is used and the corrected local track parameters and associated covariance are calculated using Eq.~\ref{eq:filterEKF}. In ACTS, the covariance in Eq.~\ref{eq:crossEKF} between the track state and the measurement is part of the covariance matrix of the local track parameters, and therefore no additional calculation of this term is needed and the Kalman gain formulas for the EKF and NLKF are identical. Therefore, the same Kalman filtering and smoothing formulae for the EKF are used for NLKF with the predicted local track parameters $x_k^{k-1}$ and its covariance $C_k^{k-1}$ in Eq.~\ref{eq:filterupdate} replaced by the corresponding corrected local track parameters.

\subsection{Comparison of the EKF and NLKF for track fitting}
\begin{figure}[htb!]
\centering
  \includegraphics[width=1.0\linewidth]{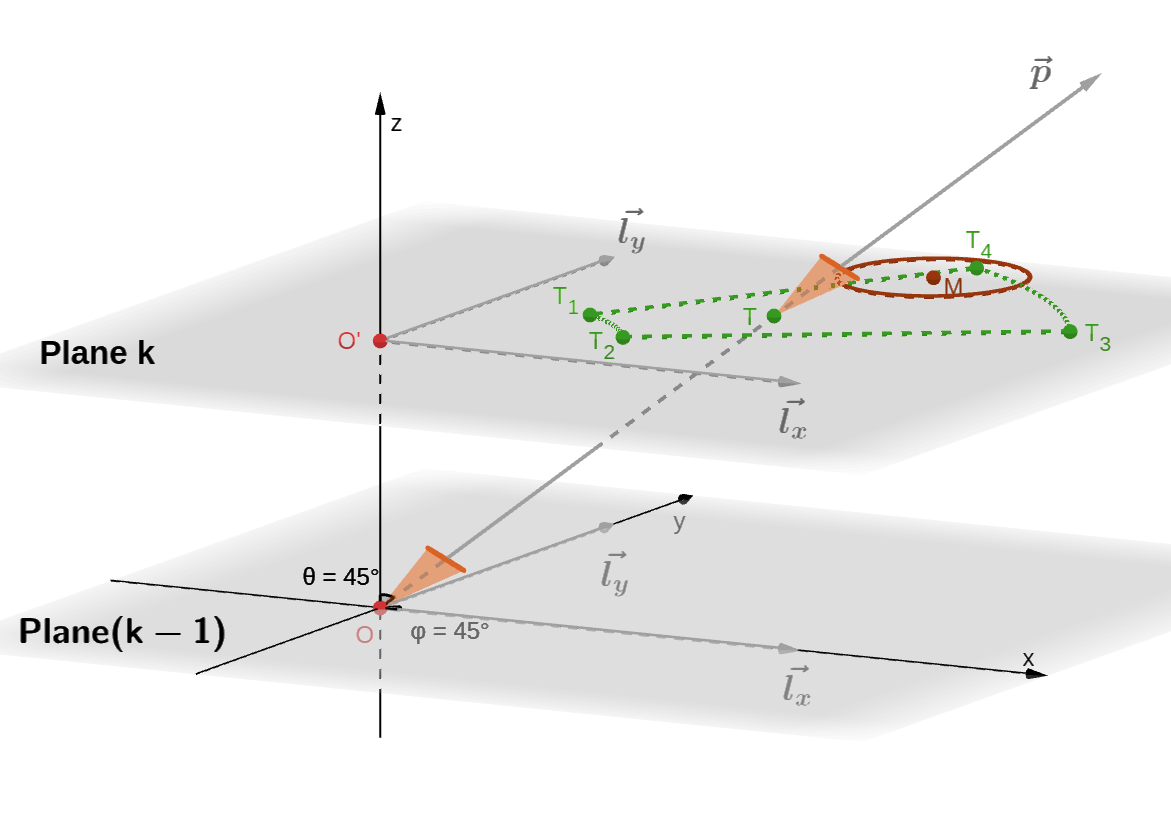}
  \includegraphics[width=1.0\linewidth]{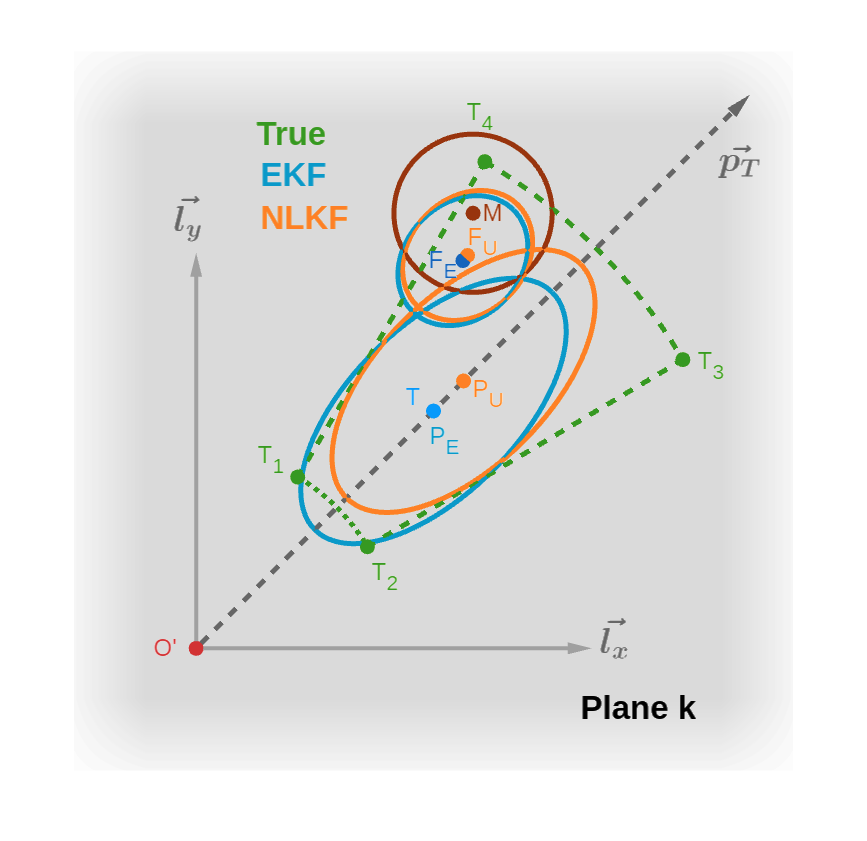}
  \caption{Illustration of the impact of non-linear effects during track parameter transformation for a two layer detector with no magnetic field. (Top) A track intersects planes $k-1$ and $k$ at points $O$ and $T$. The local coordinates of point $O$ has zero covariance. The track direction has the azimuthal and polar angles of $\pi/4$ and a covariance denoted by the orange cones. (Bottom) The true covariance of the predicted local coordinates on plane $k$ is shown by the green dashed curve, a measurement at $M$ with its covariance is denoted by the red circle, and the predicted (filtered) local coordinates of the track using the EKF and the NLKF are shown with the ellipses centered $P_E$ ($F_E$) and $P_U$ ($F_U$), respectively.}
  \label{fig:track_extra}
\end{figure}

Figure.~\ref{fig:track_extra} illustrates the impact of the non-linearity effects on track parameter propagation using the configuration shown in  Figure~\ref{fig:locals_vs_angles}. Given the configuration of the track parameters at plane $k-1$, the local coordinate of the intersection of the track on plane $k$ will have a true covariance indicated by the dashed green shape. The two arc sides, and two radial sides of the true covariance are due to variations in the polar and azimuthal angles of the track direction. 
If the EKF is used to transport the track from plane $k-1$ to plane $k$, the local coordinate of the track on plane $k$ will have the error denoted by the blue ellipse. If the NLKF, is used, the local coordinate of the track on plane $k$ will have the error denoted by the orange ellipse.

\begin{figure}[htb!]
\centering
  \includegraphics[width=1.0\linewidth]{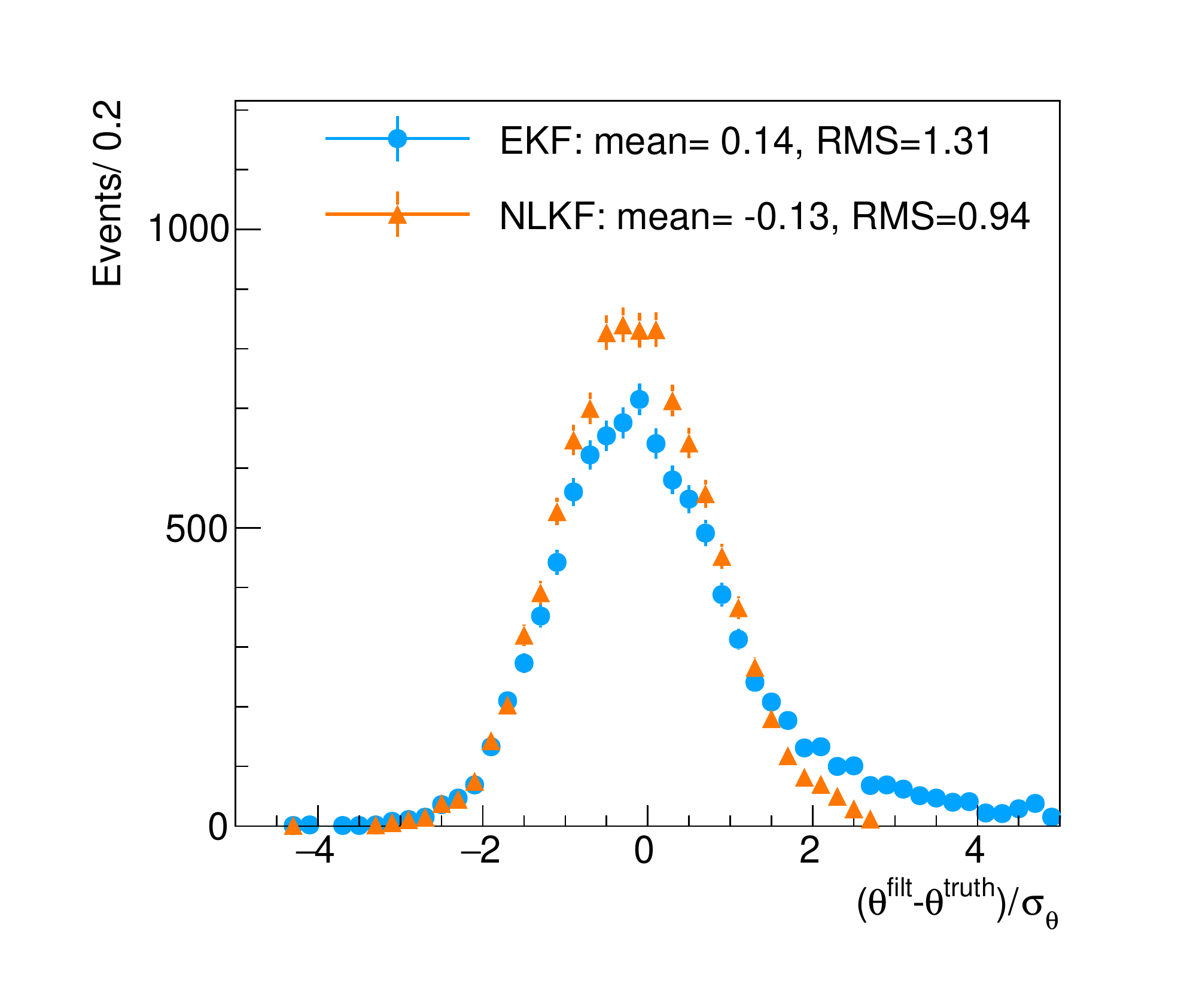}
  \caption{Comparison of pull of the filtered momentum direction polar angle $\theta$ using the EKF (blue) and NLKF (orange) with the setup in Fig.~\ref{fig:track_extra}. Ten thousand tracks and their corresponding measurements are simulated.}
  \label{fig:pull_angles_filtered}
\end{figure}

Such non-linear effects will impact the Kalman filtering procedure. In particular, the Kalman gain matrix in Eq.~\ref{eq:filtergain} tends to either over- or under- estimate the polar angle of the track. This effect is demonstrated in Fig.~\ref{fig:pull_angles_filtered} using the configuration from Fig.~\ref{fig:track_extra} and showing the pull distribution of the filtered polar angle. The pull value for track parameter $v$ is defined as,
\begin{equation}
pull_{v} = \frac{v^{fit} - v^{truth}}{\sigma_{v}}.
\label{eq:pull}
\end{equation}
Here $v^{fit}$ and $\sigma_{v}$ are the estimated value and uncertainty of the track parameter $v$ respectively, and $v^{truth}$ is the true simulated value of the $v$.
If both the values and uncertainties of the track parameters are estimated correctly, the pull distributions are expected to follow normal distributions. To avoid any bias from assuming a Gaussian distribution, the mean and the \emph{Root-Mean-Square} (RMS) values of the pulls are compared between the EKF and the NLKF.
For the EKF, the filtered polar angles are biased to larger values than their true values with a large RMS. For the NLKF, the mean of the polar angles is biased to negative values, but the RMS is significantly improved. The impact of the non-linear effects on the azimuthal angle is smaller. Both implementations have mean at zero and RMS at 1.2. 

\section{Performance studies}
\label{sec:perf}

\begin{figure}[htb!]
\centering
  \includegraphics[width=1.0\linewidth]{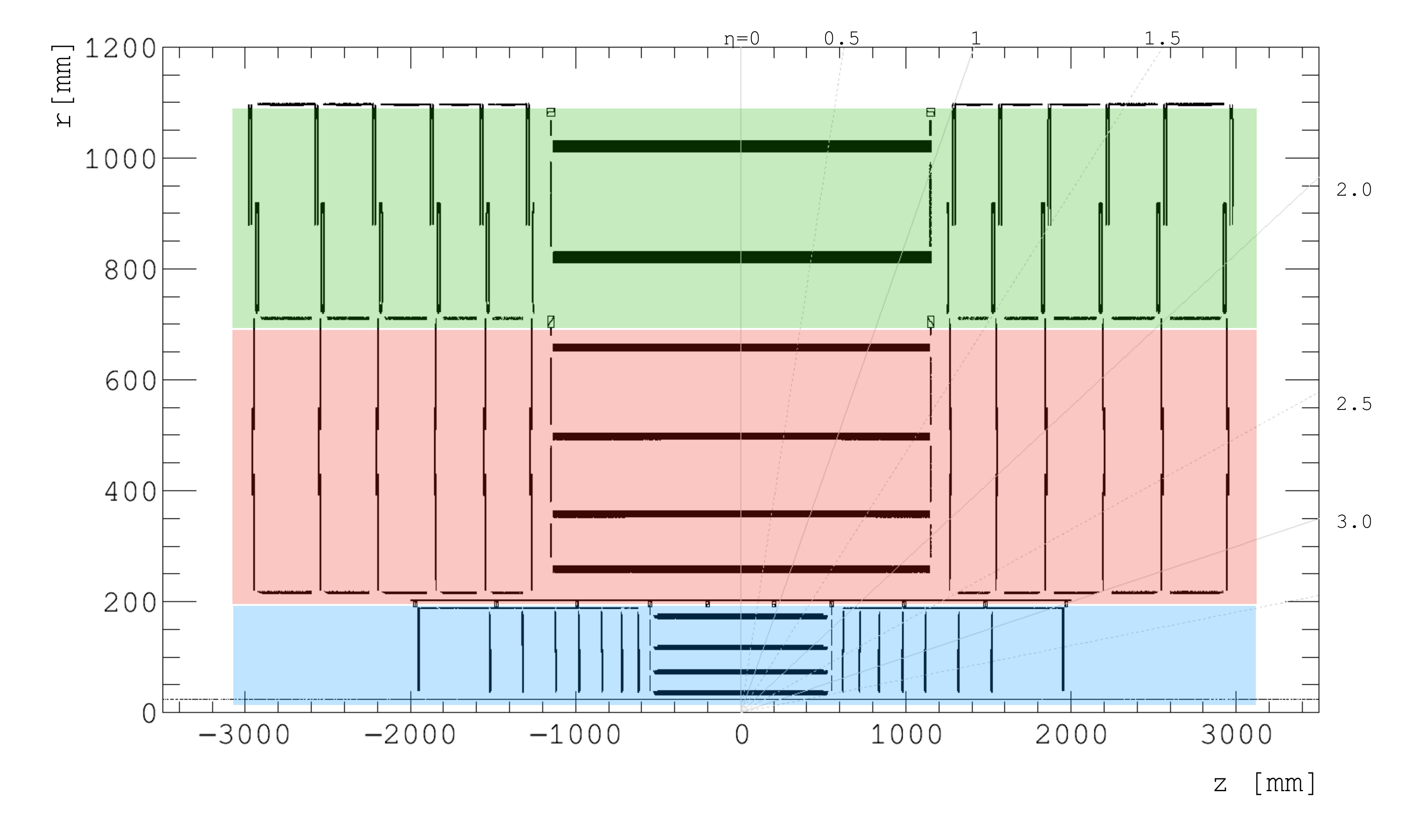}
  \caption{Schematic layout of the ODD silicon tracking detector projected into the $z$-$r$ plane. The beam interaction would occur at $z=0, r=0$. The location of the pixel detector is shown in blue, the two strip detectors with different intrinsic resolution are shown in red and green, where the inner strip detector (red) has better resolution than the outer strip detector (green).}
  \label{fig:ODD_layout}
\end{figure}

The performance of EKF is evaluated using the Open Data Detector (ODD)~\cite{ODD}. The layout of the ODD is shown in Fig.~\ref{fig:ODD_layout}. It consists of a pixel detector and two strip detectors with differing instrinsic resolution and it uses a realistic material model using the DD4hep~\cite{Petri__2017} detector description tool. The ODD is immersed in a solenoidal magnetic field of 2\,Tesla centered on the beam line. 

A sample of 1 million simulated muons is used to study the performance, as muons are insensitive to the detector material. The muons are generated with transverse momentum\footnote{Transverse momentum is the momentum in the transverse $x$-$y$ plane, $p_T = \sqrt{p_x^2+p_y^2}$} $p_T$ uniformly distributed in the range of $0.4  <p_T < 100$\,GeV and pseudorapidity\footnote{Pseudorapidity is an angular quantity calculated from the polar angle $\theta$ as follows $\eta = -\ln\left[\tan\left(\frac \theta 2\right)\right]$. $\eta = +\infty$ corresponds to the direction of the beam.} $\eta$ uniformly distributed in the range of $|\eta|<$ 3.0. The range in $p_T$ allows us to study the impact of multiple scattering, which varies with $p_t$ and the range in $\eta$ allows us to study muons that intersect the detector modules at a range of angles. The intersection points of the muons with the detectors, the simulated hits, are generated with the Fatras fast simulation engine within the ACTS toolkit. The input measurements to the Kalman filter algorithm are created by applying Gaussian smearing to the positions of the simulated hits to emulated the impact of detector resolution. One- and two-dimensional measurements in the local coordinate frames of the detector planes are created in the strip and pixel detectors of ODD, respectively, by smearing with Gaussian distributions with zero mean and different width ($\sigma$) as in Table~\ref{tab:meas_smear_sigma}.

The reconstructed seed of the track fit is emulated by smearing the vertex position, momentum and time of the generated muons using Gaussian distributions with zero mean and either momentum-dependent or constant width. The production vertex is smeared to obtain the local coordinates $d_0$ and $z_0$ using Gaussian distributions with
$\sigma = a_0 + a_1e^{-a_2p_T}$, $q/p$ is smeared using a Gaussian distribution with $\sigma = a_0/p$, and $\phi$, $\theta$ and $t$ are smeared using a Gaussian distribution with constant $\sigma$. Table~\ref{tab:seed_smear_sigma} provides the parameters used to construct the width of the Gaussian used for the smearing, which are of similar order to the resolution of the tracking detectors at current nuclear and particle physics experiments.

\begin{table}[htbp!]
\centering
\begin{tabular}{cccc}
\hline
Subdetectors &$\sigma_x$ [$\mu m$] &$\sigma_y$ [$\mu m$]\\
\hline
Pixel        &15   &15  \\
Inner strip  &43   &-   \\
Outer strip  &72   &-   \\
\hline
\end{tabular}
\caption{The width of Gaussian (with zero mean) used to smear the $x$ and $y$ coordinates of the truth hits in different sub-detectors of the ODD.
\label{tab:meas_smear_sigma}}
\end{table}

\begin{table}[htbp!]
\centering
\begin{tabular}{cl}
\hline
Track parameters   &Smearing parameters\\
\hline
\multirow{3}{3em}{$d_0$, $z_0$} &$a_0$ = 20 $\mu$m \\
                             &$a_1$ = 30 $\mu$m \\
                             &$a_2$ = 0.3 GeV$^{-1}$ \\
$\phi$, $\theta$             &$\sigma$ = $1^{\circ}$ \\
$q/p$                        &$a_0$ = 0.01 GeV$^{-1}$\\
$t$                          &$\sigma$ =1 ns \\
\hline
\end{tabular}
\caption{The parameters used used to construct the width of the Gaussian for smearing the generated vertex, momentum and $t$ to obtain the seed of the track fit.
\label{tab:seed_smear_sigma}}
\end{table}

The physics and the computational performance of the EKF and NLKF are studied. The non-linear correction for the NLKF is only performed when the incidence angle of a track with a detector plane is larger than 0.1. The NLKF performance is found to be insensitive to the tuning parameter $\alpha$ so a fixed value of $\alpha=0.1$ is used. 

\subsection{Track parameter estimation}

The mean and RMS of the residuals, defined as $v^{fit} -v^{truth}$, and the pulls, defined in  Eq.~\ref{eq:pull}, of the perigee track parameters are used to evaluate the performance. The pull depends on the central value of the track parameter and its uncertainty, but the residual depends only on the central value. Ideally, the pulls would have means of zero and and RMSs of one and the residuals would have means of zero and the RMS of the detector resolution.

The mean and the RMS of the residuals and pulls are studied in bins as a function of $\eta$.  
The degree of non-linear effects, the number of detector layers and the amount of material that a charged particle passes through vary with $\eta$. The results are presented with and without a magnetic field and with and without the impact of the particle interactions with the detector material.
The results without the magnetic field are equivalent to a track fitting scenario without non-linear effects in the track state propagation model in which case non-linear effects are only due to the measurement model, which is directly addressed by this implementation. The impact of the non-linear effects on $t$ is negligible and therefore only the RMS of its pull as a function of $\eta$ is shown. 

\begin{figure*}[hbtp!]
\centering
\includegraphics[width=0.32\linewidth]{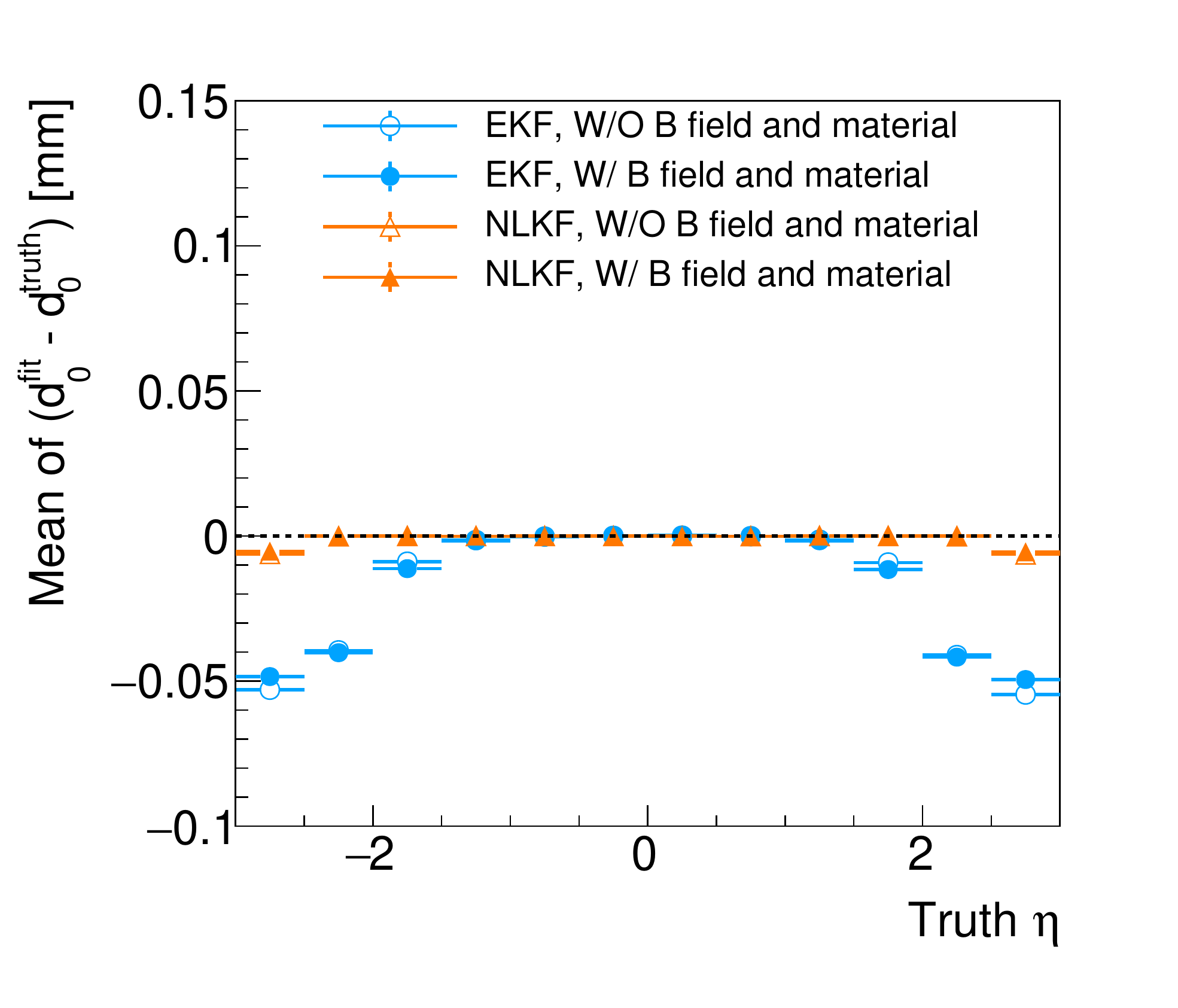}
\includegraphics[width=0.32\linewidth]{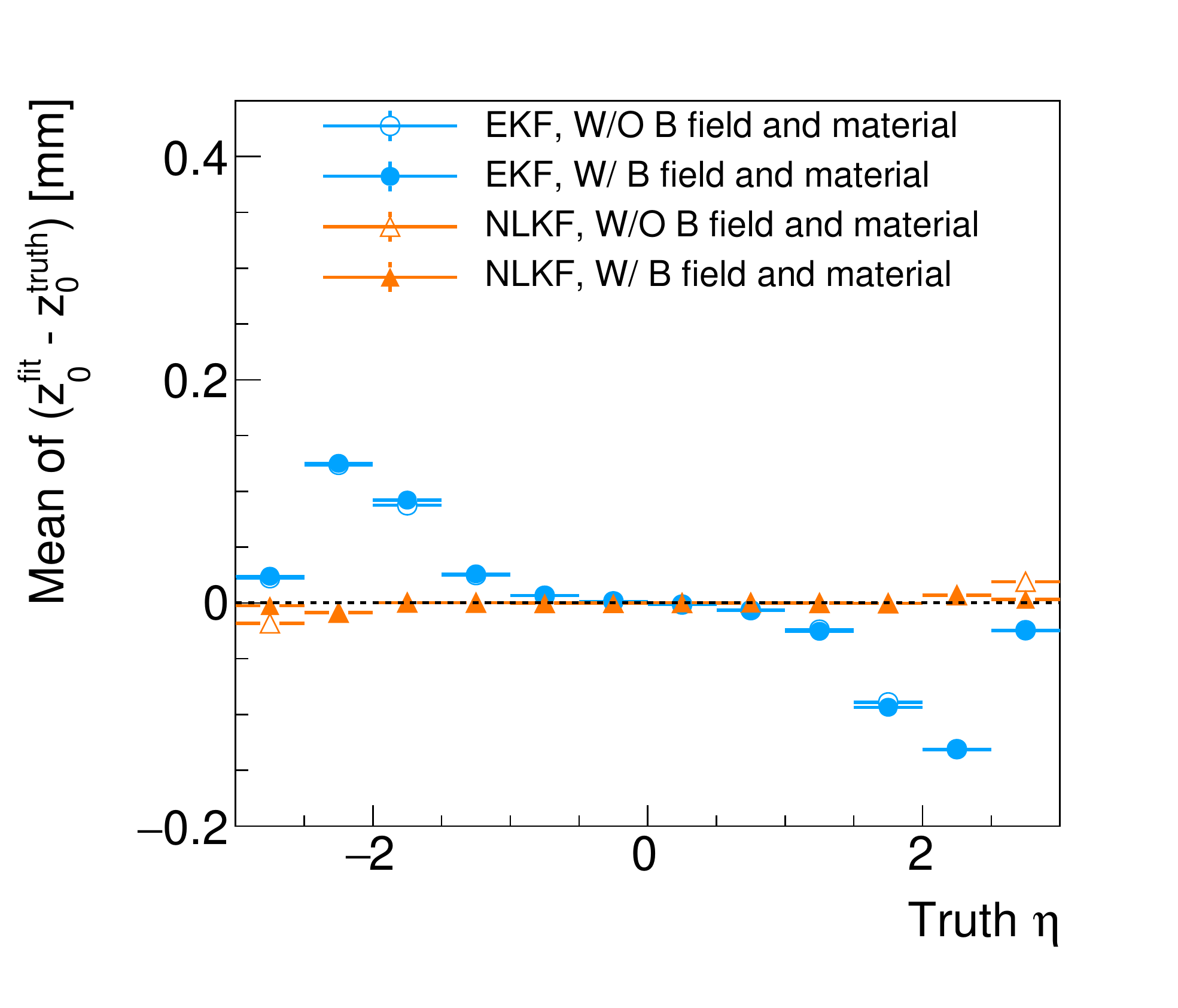}\\
\includegraphics[width=0.32\linewidth]{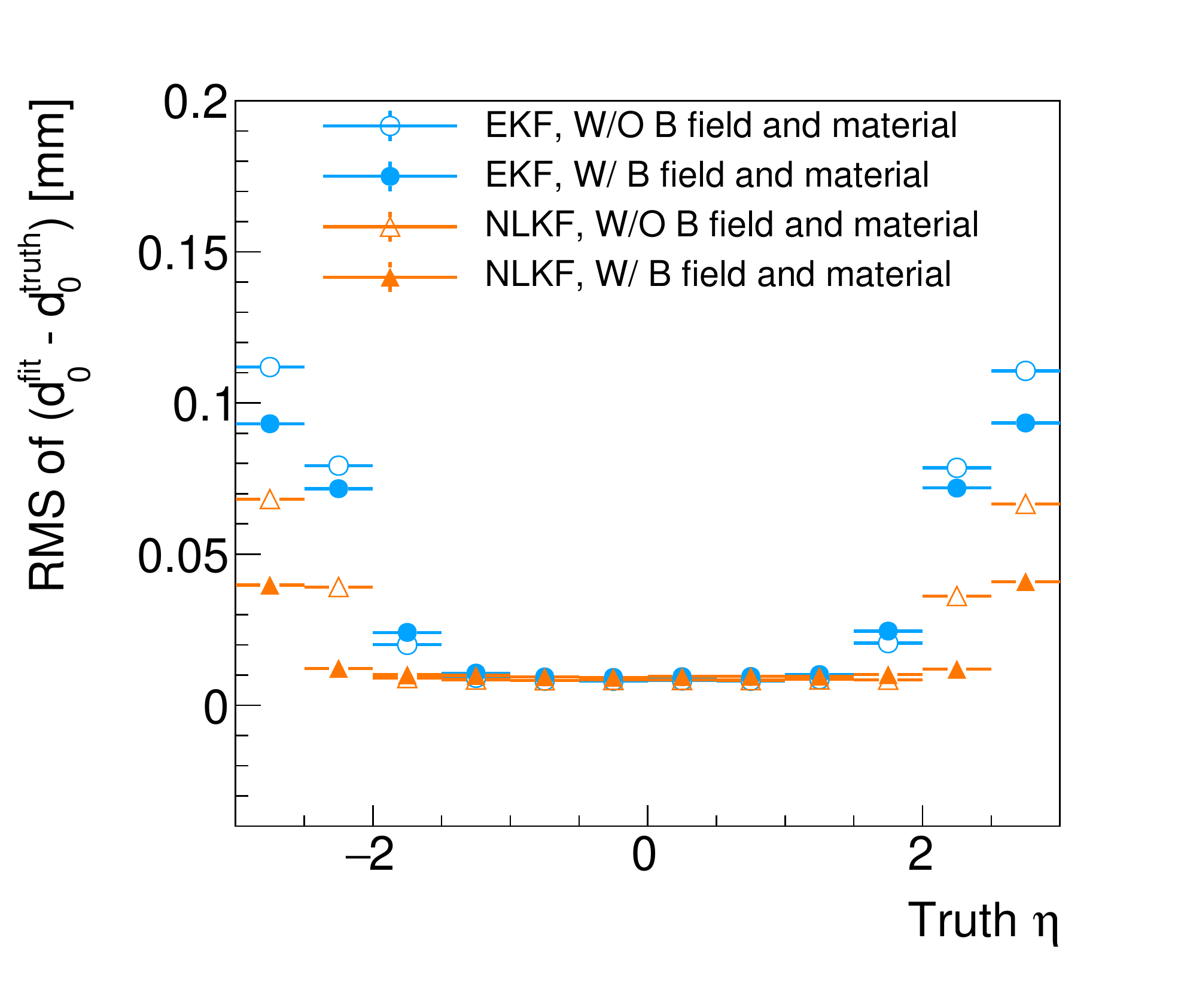}
\includegraphics[width=0.32\linewidth]{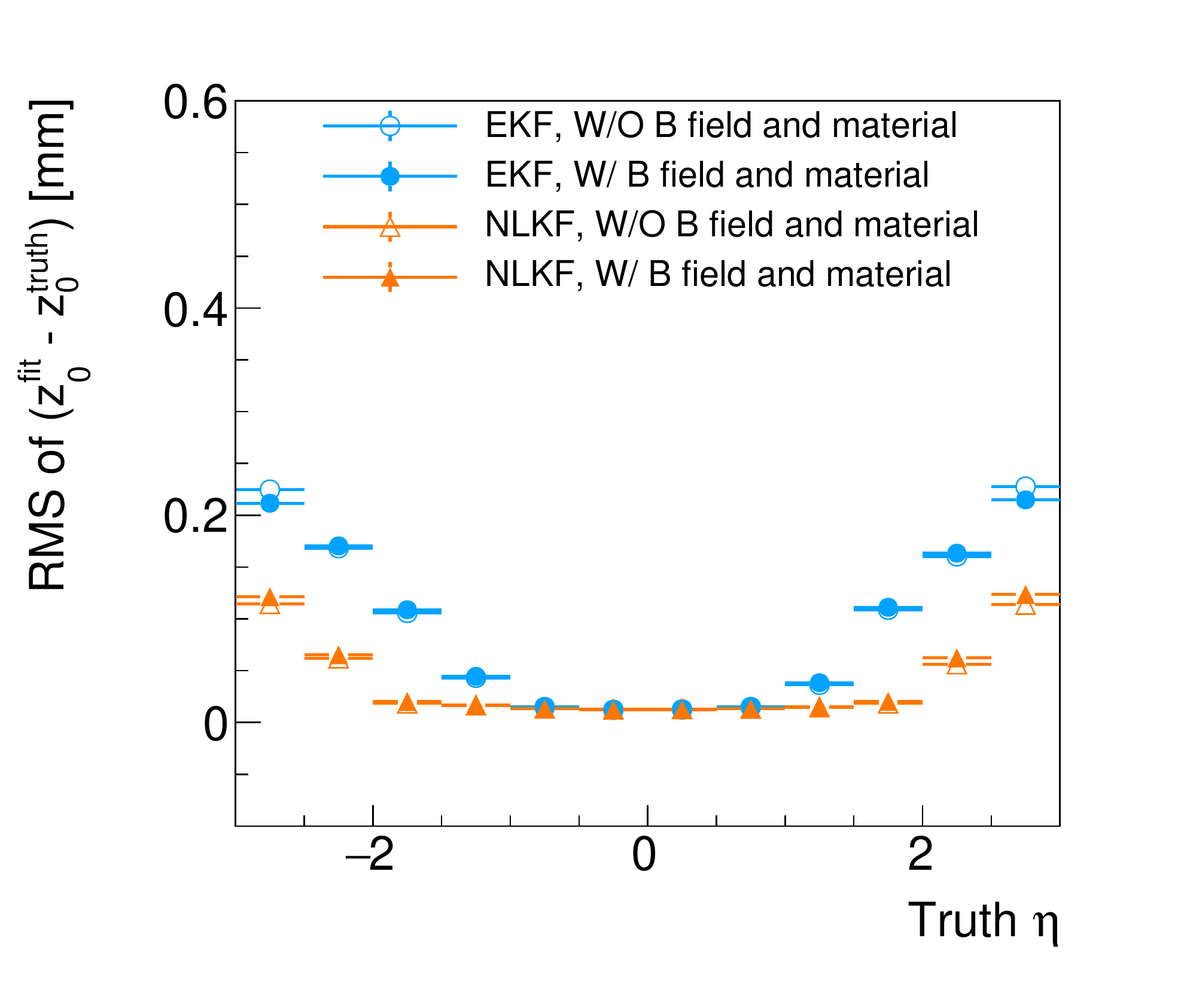}
\caption{The mean (top) and RMS (bottom) of the residual of fitted perigee track parameters $d_0$ (left) and $z_0$ (right) parameterized as a function of simulated particle $\eta$ (20\,$<p_T<$\,100\,GeV) for the ODD without (blue circles for EKF, orange hollow triangles for NLKF) and with (blue dots for EKF, orange filled triangles for EKF ) the presence of a solenoidal magnetic field of 2\,T and material effects. The dashed horizontal lines in the upper panel denote the expected mean of the residuals. \label{fig:ODD_ATLASField_resMeanRMS_vs_eta}}
\end{figure*}

 The mean of the residuals of the impact parameters, $d_0$ and $z_0$, as a function of $\eta$ for simulated particles with $p_T>20$\,GeV are shown in the upper panel of Fig.~\ref{fig:ODD_ATLASField_resMeanRMS_vs_eta}. The mean estimated using the EKF is biased from zero at higher $|\eta|$ bins due to more pronounced non-linear effects in this region. However, such biases are absent when the NLKF is used. As there is a strong correlation between the residuals and pulls, the mean of the pulls show similar biases to the residual means of the perigee track parameters.

The resolution of the impact parameters as a function of $\eta$ for simulated particles with $p_T>$\,20\,GeV are shown in the lower panel of Fig.~\ref{fig:ODD_ATLASField_resMeanRMS_vs_eta}. The NLKF improves their resolution by up to 50\% at higher $|\eta|$ bins compared to the EKF. There is similar improvement for $\phi$ and $\theta$. All track parameters are studied, and no improvement in the resolution of $q/p$ is observed.

\begin{figure*}[!htbp]
\centering
\includegraphics[width=0.32\linewidth]{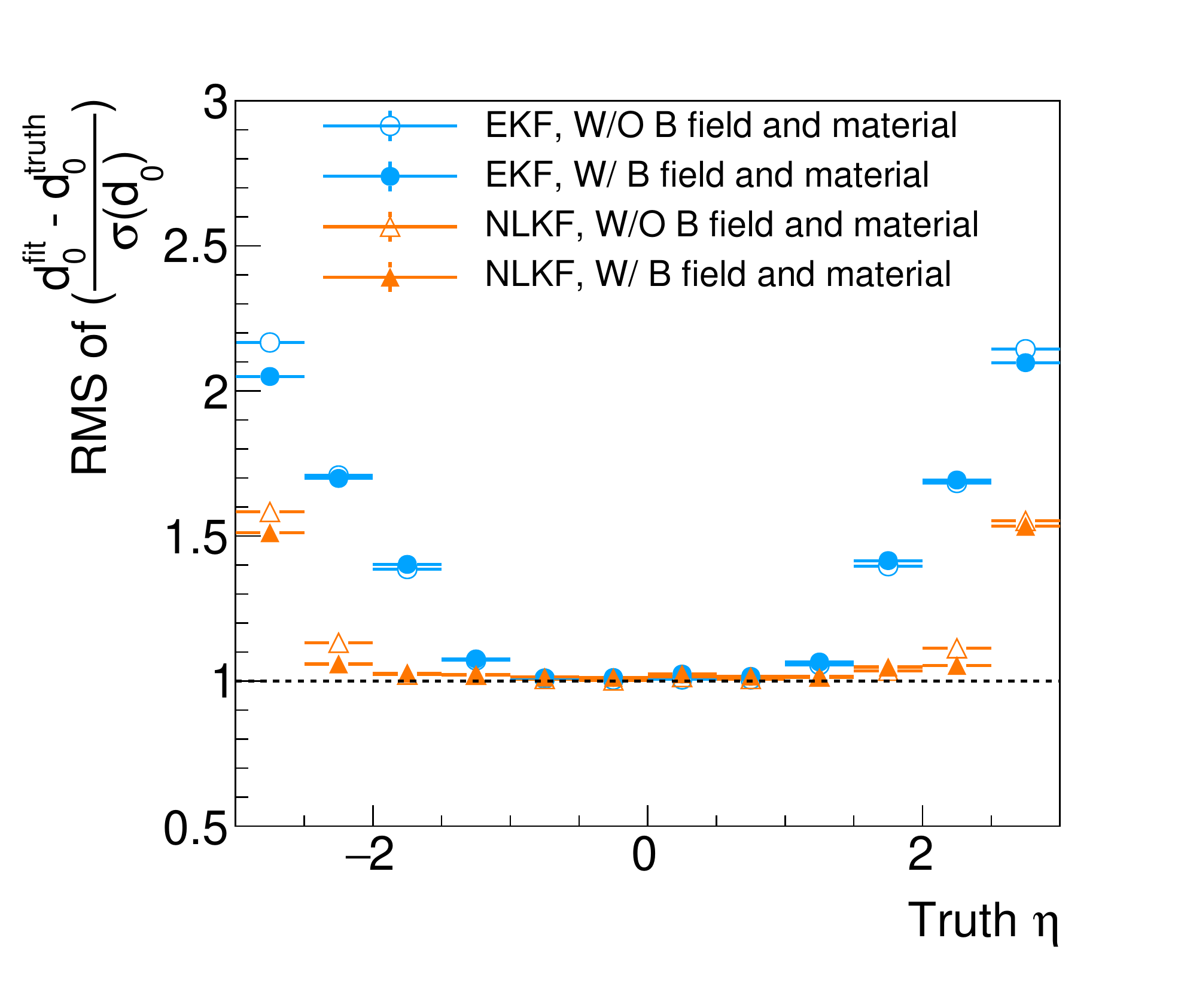}
\includegraphics[width=0.32\linewidth]{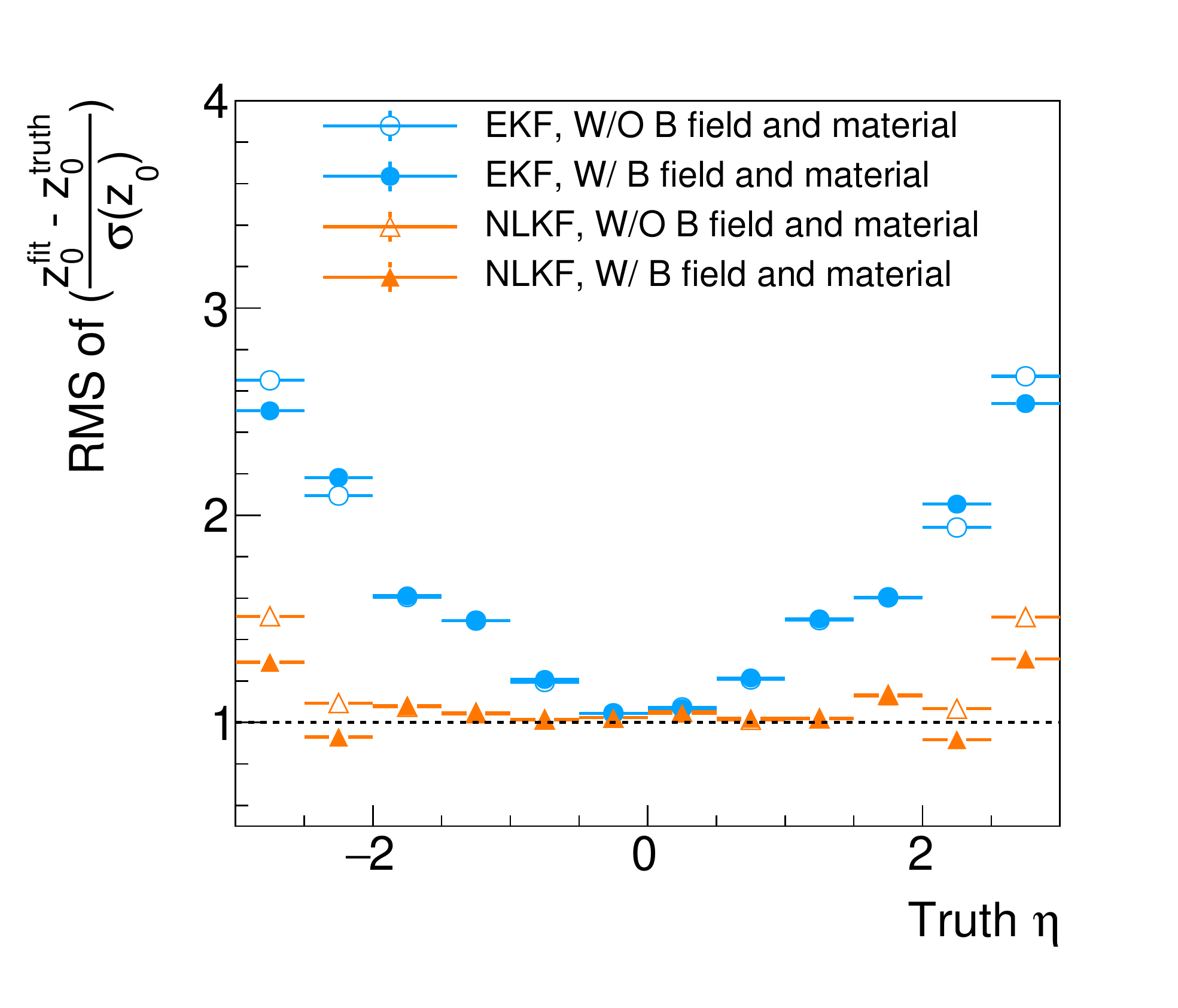}
\includegraphics[width=0.32\linewidth]{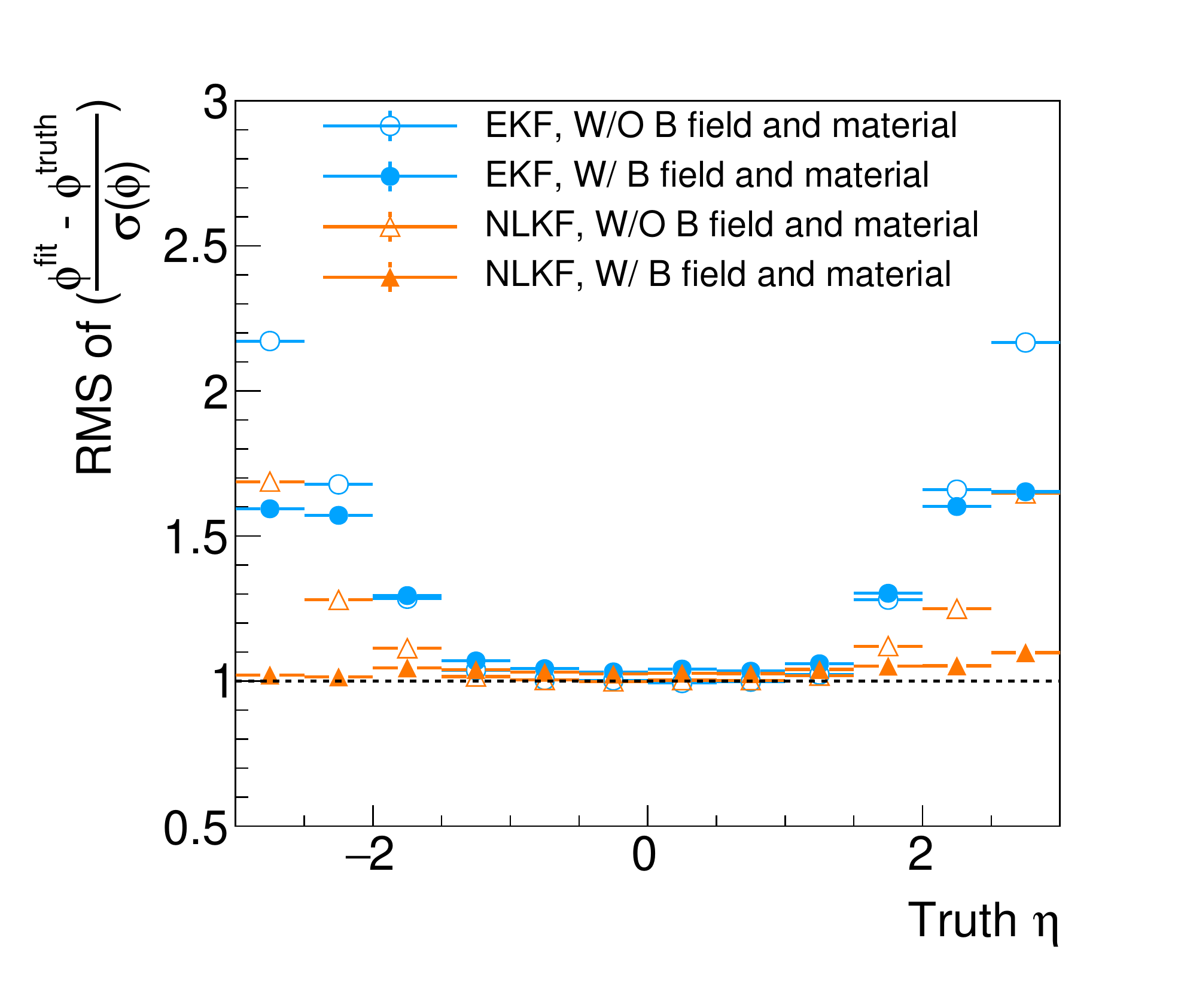}
\includegraphics[width=0.32\linewidth]{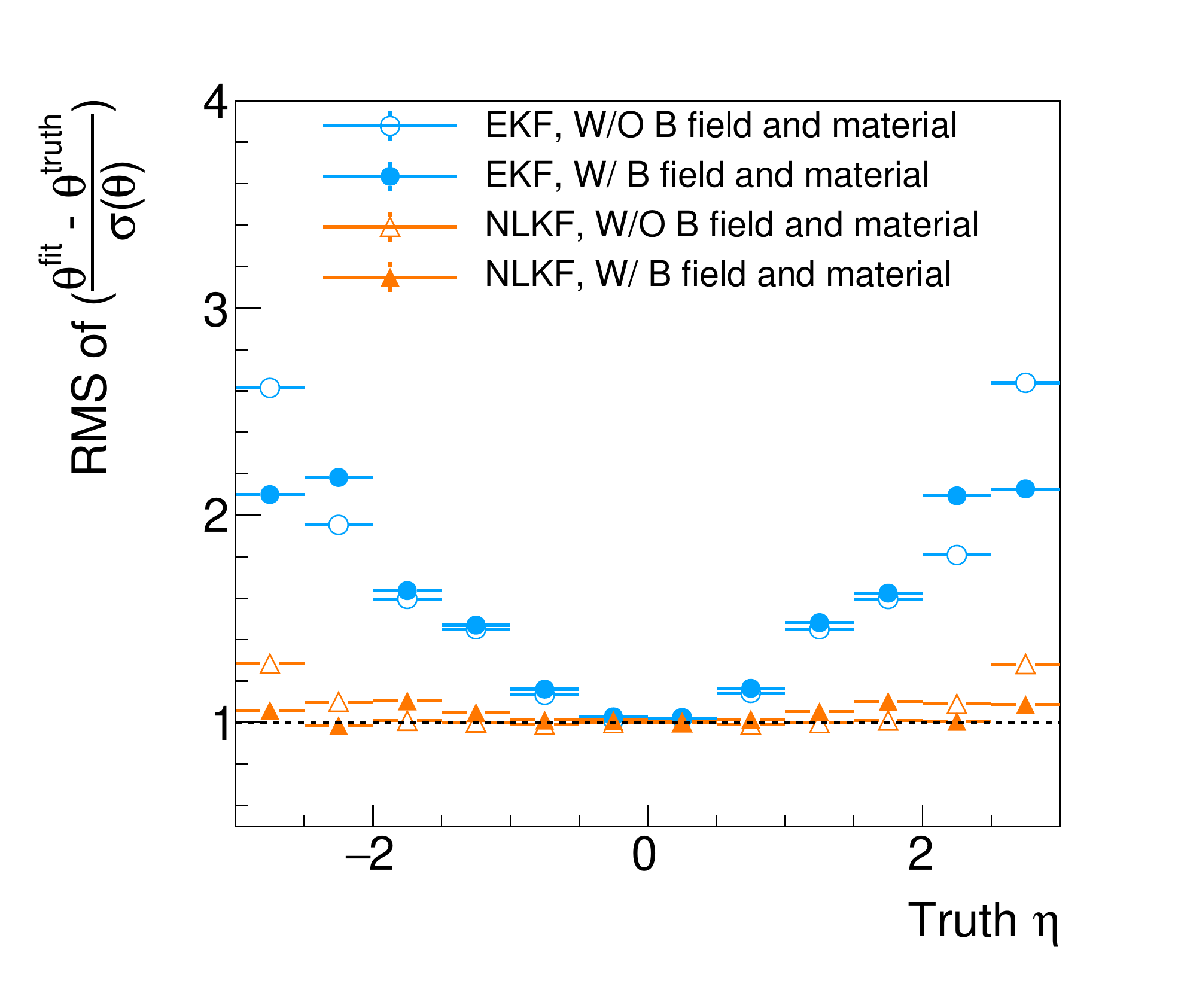}
\includegraphics[width=0.32\linewidth]{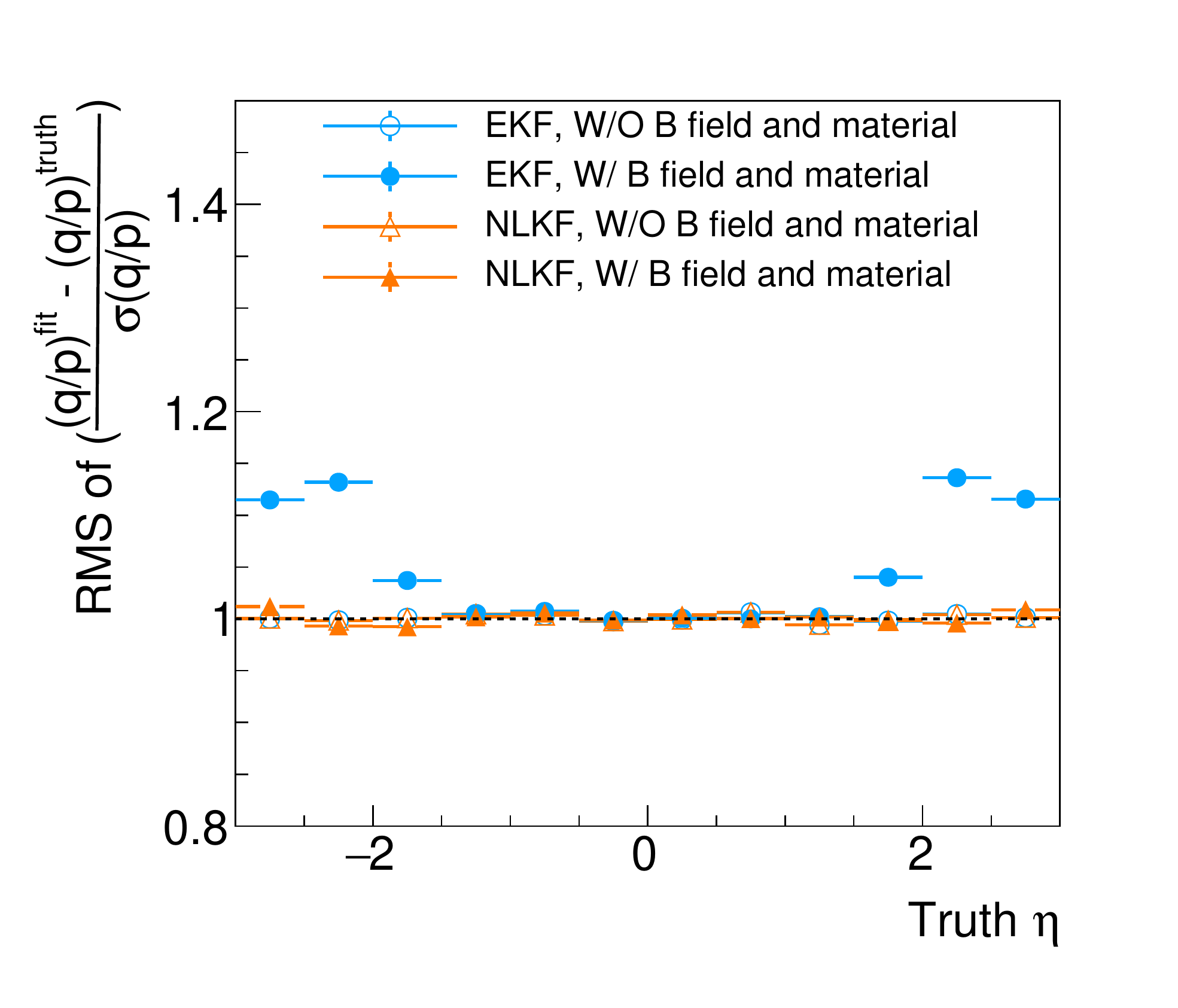}
\includegraphics[width=0.32\linewidth]{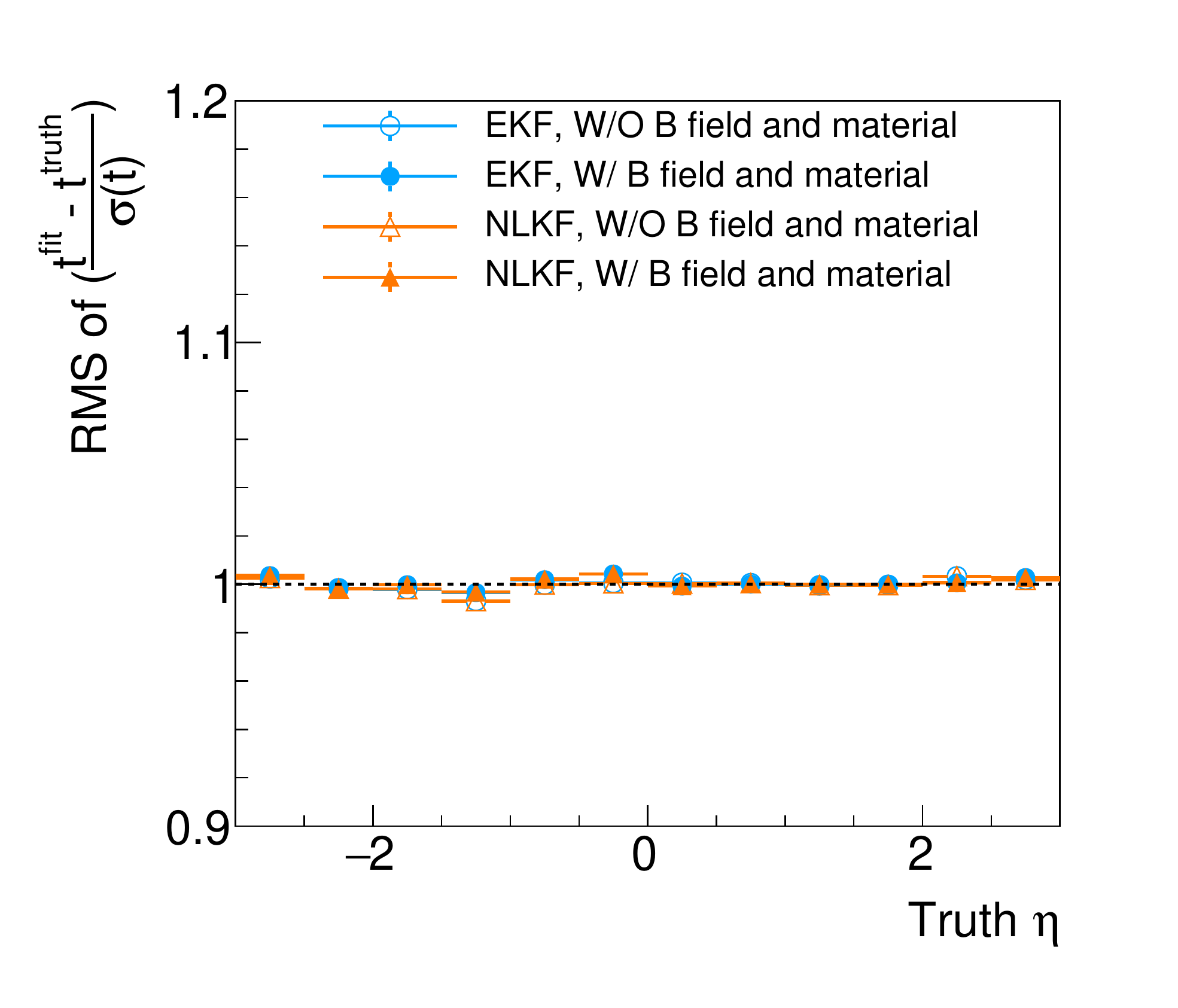}
\caption{The RMS of the pull of fitted perigee track parameters $d_0$, $z_0$, $\phi$, $\theta$, $q/p$ and $t$ parameterized as a function of the simulated particle $\eta$ ($20<p_T<100$\,GeV) for the ODD without (blue circles for EKF, orange hollow triangles for NLKF) and with (blue dots for EKF, orange filled triangles for EKF ) the presence of a solenoidal magnetic field of 2\,T and material effects. The dashed horizontal lines denote the expected RMS of the pulls. \label{fig:ODD_ATLASField_pullRMS_vs_eta}}
\end{figure*}

 Fig.~\ref{fig:ODD_ATLASField_pullRMS_vs_eta} shows the RMS of the pulls of all perigee track parameters as a function of $\eta$ for simulated particles with $p_T>$\,20\,GeV. The parameter $t$ is unaffected by the non-linear effects and hence the RMS of its pulls is approximately one. Non-linear effects cause the RMS to deviate from one at higher $|\eta|$ for $d_0$, $z_0$, $\phi$, $\theta$ and $q/p$ when using the EKF. The deviation is largest for $z_0$ and $\theta$ where the RMS can reach up to 2.6 and smallest for $q/p$. The deviation is significantly reduced using the NLKF, i.e.~the RMS for all track parameters is below 1.7 in the whole $\eta$ range being studied and below 1.3 in the central region. No deviation for $\phi$, $\theta$ and $q/p$ is observed with the NLKF for tracks in such $p_T$ range when magnetic field and material effects are present.
 
\begin{figure*}[!htbp]
\centering
\includegraphics[width=0.32\linewidth]{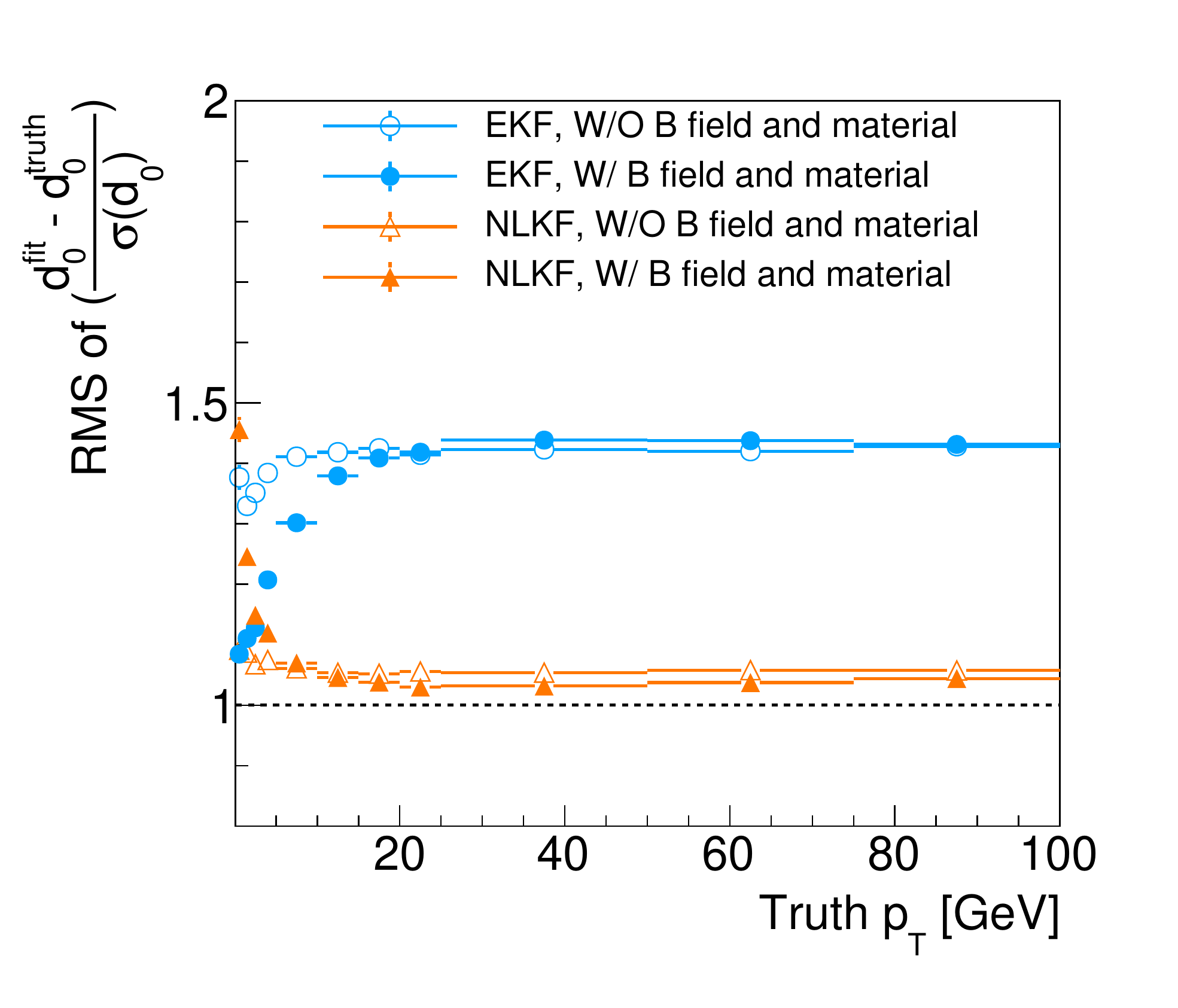}
\includegraphics[width=0.32\linewidth]{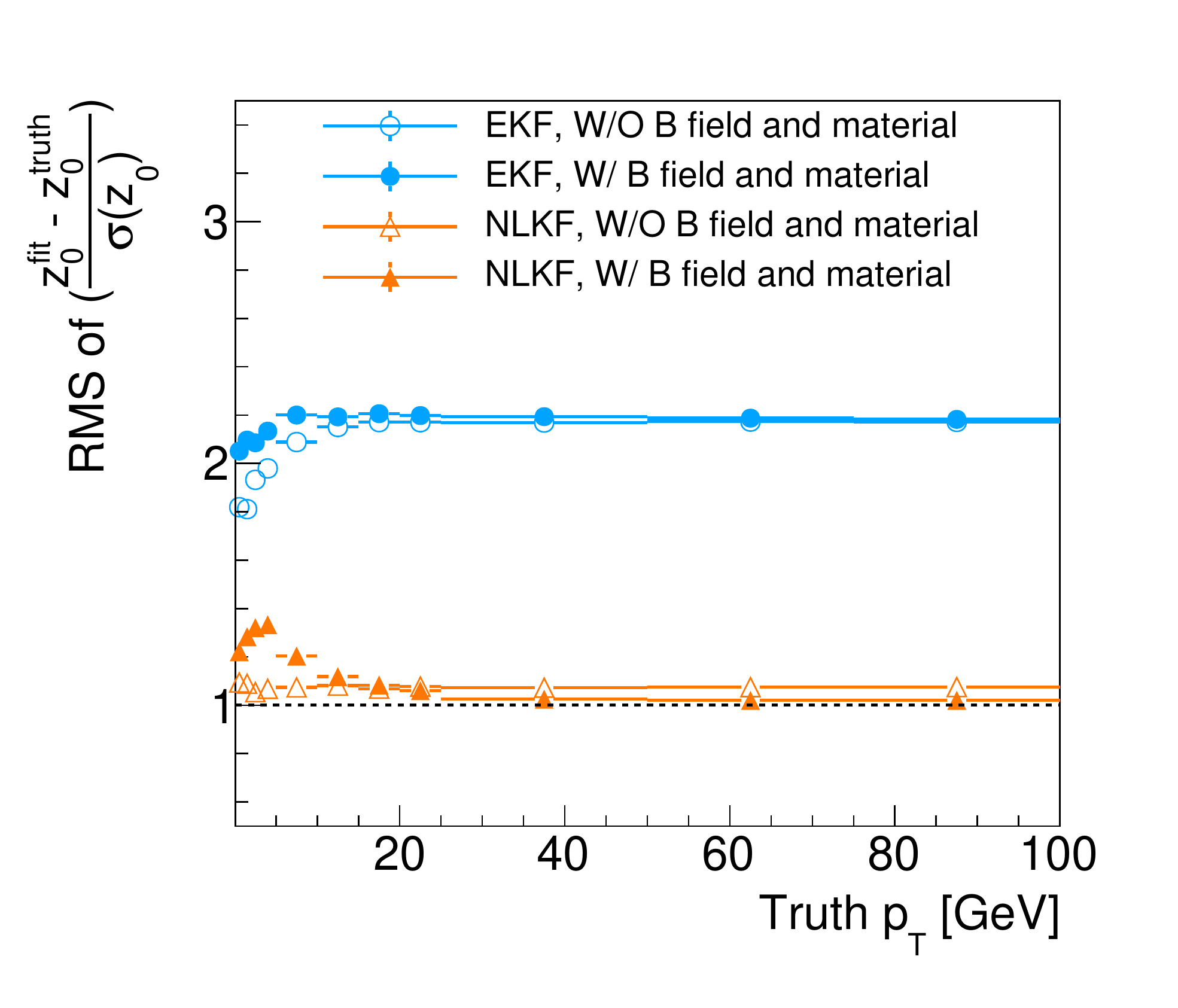}
\caption{The RMS of the pull of fitted perigee track parameters $d_0$ and $z_0$ parameterized as a function of simulated particle $p_T$ ($1.0<|\eta|<2.5$) for the ODD without (blue circles for EKF, orange hollow triangles for NLKF) and with (blue dots for EKF, orange filled triangles for EKF ) the presence of a solenoidal magnetic field of 2\,T and material effects for the track parameters $d_0$, $z_0$, $\phi$, $\theta$ and $q/p$, respectively. The dashed horizontal lines denote the expected RMS of the pulls. \label{fig:ODD_ATLASField_pullRMS_vs_pt}}
\end{figure*}

The dependence of the pulls on track $p_T$ is studied in Fig.~\ref{fig:ODD_ATLASField_pullRMS_vs_pt}, which shows the RMS of the pulls of the impact parameters for simulated particles in the range of $1.0<|\eta|<2.5$. This $\eta$ range was selected because non-linear effects are significant for these values. The RMS of the pulls is smaller for lower $p_T$ tracks using the EKF. When there is no material and magnetic field, the deviation with the NLKF is always significantly smaller than that with the EKF for all values of $p_T$.  
When there is material and magnetic field, the NLKF achieves significantly better performance than the EKF for $z_0$. For $d_0$, the NLKF improves the RMS of the pulls for tracks with $p_T>2$\,GeV despite the fact that it corrects the bias of the mean of residual and pull for tracks in all values of $p_T$. The performance differences observed between the NLKF and the EKF for $\phi$ are similar to $d_0$ and for $\theta$ are similar to $z_0$. This is expected due to the correlations between the pairs of track parameters.

\subsection{Computational Performance}
Additional computational cost with the NLKF is expected due to the additional evaluation points, which are key to improving the precision. An estimate of this cost is obtained by comparing the track fitting time of the NLKF to that of the EKF as a function of $\eta$ and $p_T$. In each $\eta$ or $p_T$ bin, track fitting is performed five times per sample with 1k tracks. The mean of the track fitting time per track from the five tests is shown as the nominal value, and the RMS is shown as the uncertainty bar. The tests are performed in a single thread using the Intel Core i7-8559U CPU @2.70 GHz processor.

\begin{figure*}[!htbp]
\centering
  \includegraphics[width=0.32\linewidth]{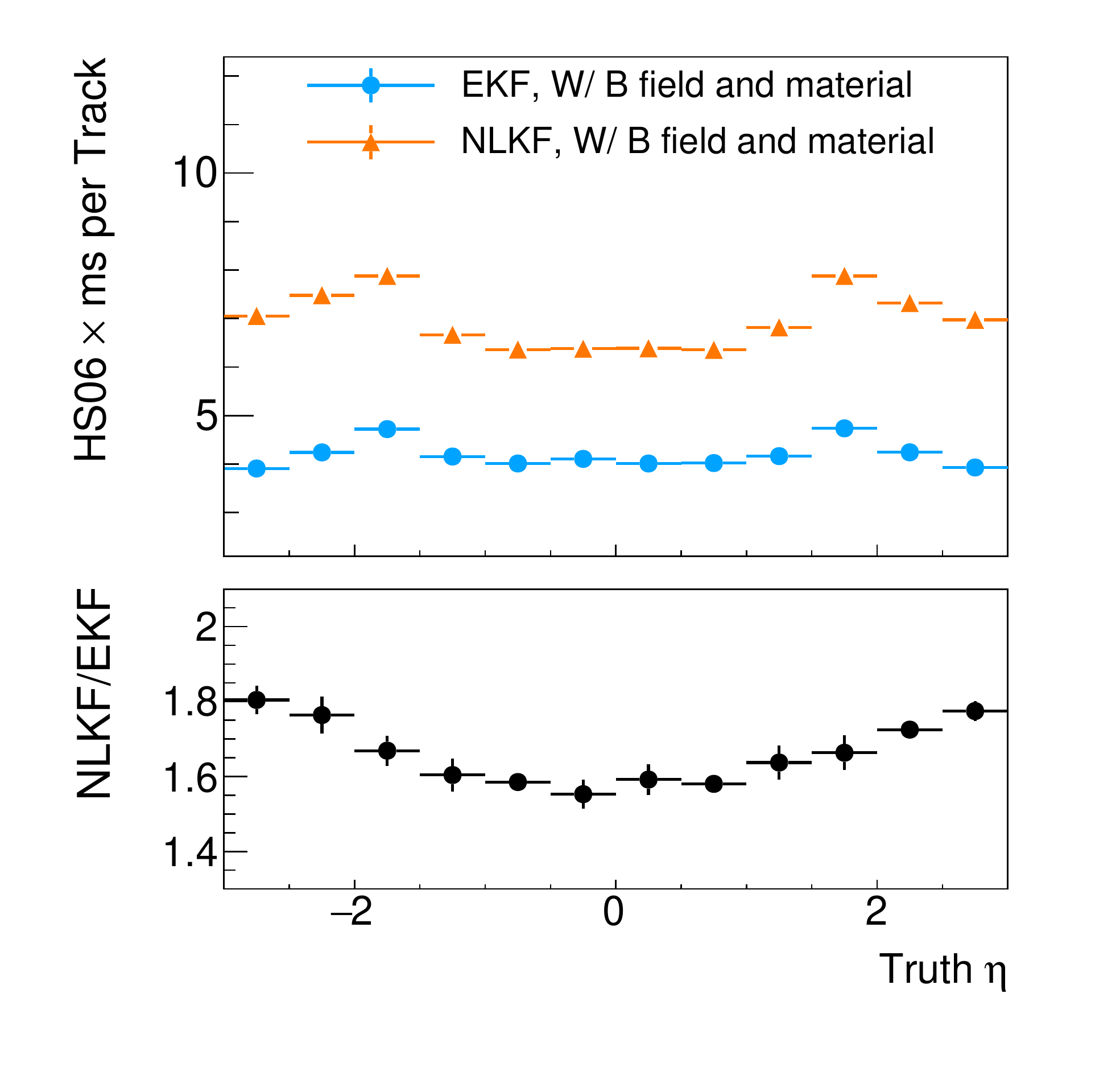}%
  \includegraphics[width=0.32\linewidth]{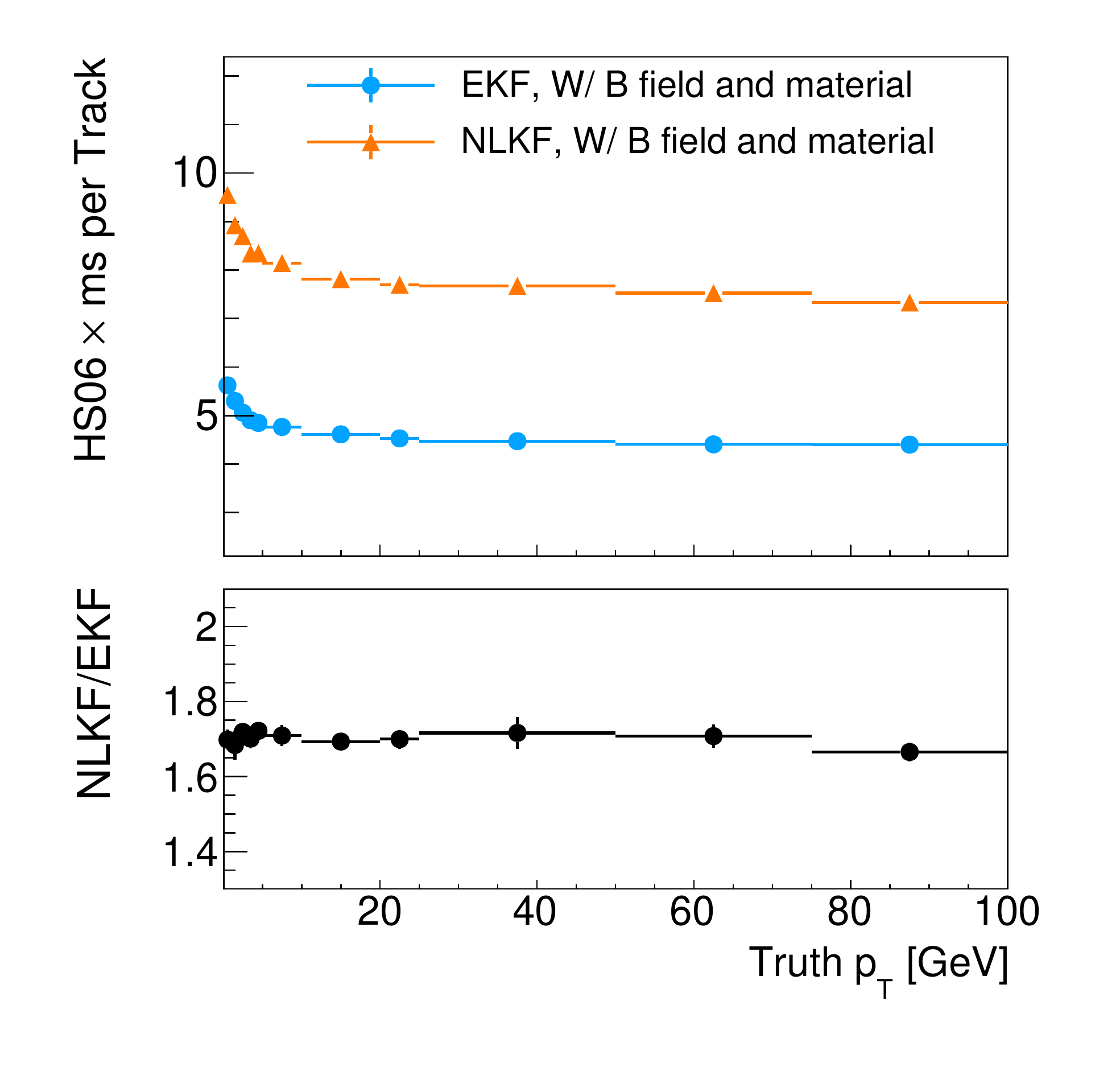}%
  \caption{(Left) A comparison of the track fitting time per track as a function of simulated particle $\eta$ ($20<p_T<100$\,GeV) between EKF and NLKF for the ODD at ATLAS B field. (Top) The track fitting time in HS06\,$\times$\,ms per track. The blue dots and orange triangles shown the results obtained using EKF and NLKF, respectively. (Bottom) The ratio of fitting time per track between NLKF and EKF. (Right) A comparison of the track fitting time per track as a function of simulated particle $p_T$ ($1.0<|\eta|<2.5$) between EKF and NLKF for the ODD at a solenoidal magnetic field of 2\,T. (Top) The track fitting time per track. The blue dots and orange triangles shown the results obtained using EKF and NLKF, respectively. (Bottom) The ratio of fitting time per track between NLKF and EKF.
  \label{fig:ODD_ATLASField_time}  }
\end{figure*}

Fig.~\ref{fig:ODD_ATLASField_time} shows the track fitting time in HS06~\cite{hepspec}\,$\times$\,ms per track as a function of $\eta$ or $p_T$ of the simulated particles with EKF and NLKF.
The average fitting time per track with EKF is approximately 4.8 HS06\,$\times$\,ms and with NLKF it increases by a factor ranging from $\sim 1.6$ in the barrel region to $\sim 1.8$ at higher $\eta$.  In general, track parameter estimation is not the most timing consuming step during track reconstruction, therefore this can be expected to have a negligible impact on the total time for track reconstruction in most applications. 

\section{Conclusion}

The reconstruction of charged particle trajectories is a challenging computational task for nuclear and particle physics experiments today and in the future. The Kalman Filter algorithm is currently widely used due to its excellent performance, however, it is limited by its assumption of linear models for the system and measurements as well as Gaussian distributions for the noise. We have introduced the non-linear Kalman filter for charged particle reconstruction, which uses a set of discretely sampled points to account for non-linear effects.

We tested the performance of our NLKF algorithm using the ODD. The NLKF yields residuals for all track parameters with a mean of zero throughout $\eta$. In addition, the RMS of the residuals are reduced for most track parameters, by up to a factor of two. The effect is most pronounced in regions with 
larger incidence angle of the tracks on the measurement planes, which are located at large values of $|\eta|$ in the detector geometry we studied. Compared to the EKF, the NLKF also provides a more accurate estimation of the uncertainty of the parameters, which results in the RMS of the pulls being more consistent with one for a larger range of $\eta$. The improvement is more pronounced for tracks with larger $p_T$.

The computational requirements for the NLKF increase due to the additional evaluation points. We found that the time for track fitting increases from a factor of 1.6 to 1.8 depending on the $p_T$ and $\eta$ of the particle. However, track fitting is typically a small fraction of the total track reconstruction time in most applications.

In conclusion, the NLKF shows promising performance in improving the estimation of the track parameters corresponding to charged particle trajectories by accounting for non-linear effects. In particular, its use can be warranted in applications where the precision of the track parameters is particularly important.

\afterpage{\clearpage}
\label{sec:concl}


\section*{Acknowledgments}
Xiaocong Ai, Nicholas Styles acknowledge support from DESY (Hamburg, Germany), a member of the Helmholtz Association HGF.

\section*{Declarations}

\paragraph{Funding}
This work was funded by the NSF under Cooperative Agreement OAC-1836650.

\paragraph{Conflict of interest}
The authors declare that they have no conflict of interest.

\paragraph{Availability of data and material} 
Not applicable. No associated data except for code.

\paragraph{Code availability}
The code used for this research is available open source~\cite{acts_on_github}.


\bibliography{biblib}

\end{document}